\documentclass[12pt]{article} 

\usepackage{epsfig}
\usepackage{graphicx}
\usepackage{psfrag}
\usepackage{amssymb}
\usepackage{amsmath}

\newcommand{\rf}[1]{(\ref{#1})}
\newcommand{\beq}{\begin{equation}}
\newcommand{\beql}[1]{\beq\label{#1}}
\newcommand{\eeq}{\end{equation}}
\newcommand{\bea}{\begin{eqnarray}}
\newcommand{\eea}{\end{eqnarray}}

\newcommand{\e}{\mbox{e}}

\newcommand{\lam}{\lambda}

\renewcommand{\a}{\alpha}
\newcommand{\n}{\nu}
\newcommand{\m}{\mu}

\newcommand{\ep}{\varepsilon}

\newcommand{\Del}{\Delta}
\newcommand{\sg}{\sigma}

\newcommand{\ra}{\rangle}
\newcommand{\la}{\langle}

\newcommand{\dphi}{\dot{\phi}}

\begin{document}

\begin{center}
\vspace{24pt}

{ \large \bf CDT Quantum Toroidal Spacetimes: An Overview}

\vspace{24pt}

{\sl J. Ambjorn}$\,^{a,c}$,
{\sl Z. Drogosz}$\,^b$
{\sl J. Gizbert-Studnicki}$\,^{b}$
{\sl A. G\"{o}rlich}$\,^{b}$,\\
{\sl J. Jurkiewicz}$\,^{b}$ and
{\sl D. N\`{e}meth}$\,^b$

\vspace{10pt}

{\small

$^a$~The Niels Bohr Institute, Copenhagen University\\
Blegdamsvej 17, DK-2100 Copenhagen \O , Denmark.

\vspace{10pt}

$^b$~Institute of Theoretical Physics, Jagiellonian University,\\
 \L ojasiewicza 11, Krak\'ow, PL 30-348, Poland.

\vspace{10pt}

$^c$~Institute for Mathematics, Astrophysics and Particle Physics
(IMAPP)\\ Radboud University Nijmegen, Heyendaalseweg 135, \\
6525 AJ  Nijmegen, The Netherlands\\

}

\end{center}

\vspace{24pt}

\begin{center}
{\bf Abstract}
\end{center}

\noindent
{Lattice formulations of  gravity can be used to study non-perturbative aspects of quantum gravity. Causal Dynamical 
Triangulations (CDT) is a lattice model of gravity that has been used in this way. It has a built-in time foliation but is coordinate-independent in the spatial directions.  The higher-order phase transitions observed in the model may be used to define a continuum
limit of the lattice theory. Some aspects of the transitions are better studied when the topology of space is toroidal rather than 
spherical. In addition, a toroidal spatial topology allows us to understand more easily the nature of typical quantum fluctuations of the 
geometry. In particular, this  topology makes it possible to use massless scalar fields that are solutions to Laplace's equation with special boundary conditions as coordinates that capture the fractal structure of the quantum geometry. When such scalar fields are included as dynamical fields in the path integral, they can have a dramatic effect on the geometry.}


\vfill

\noindent 
----------------------------------------------------------------------------------

\noindent
{\footnotesize 
\begin{tabbing}
email:~ \= ambjorn@nbi.dk,~zbigniew.drogosz@doctoral.uj.edu.pl,\\
~ \> jakub.gizbert-studnicki@uj.edu.pl,~andrzej.goerlich@uj.edu.pl,\\
 ~\>jerzy.jurkiewicz@uj.edu.pl,~nemeth.daniel.1992@gmail.com. 
 \end{tabbing}}

\newpage


\section{Introduction}
\label{intro}

Causal Dynamical Triangulations (CDT) is an attempt to formulate a non-per\-turbative  lattice theory of quantum gravity (see 
\cite{physrep,loll} for reviews). 
As in other lattice field theories, the length of the links $\ep$ provides us with an UV cut-off, 
and the continuum limit should be obtained by removing the cut-off, i.e., by adjusting the bare (dimensionless) lattice 
coupling constants in such a way that $\ep \to 0$ while some suitably chosen physical observables remain constant 
\cite{munster}.
The use of so-called ``dynamical triangulations'' (DT) goes all the way back to attempts to provide a non-perturbative 
formulation of Polyakov's string theory. It provided a lattice formulation where the functional integration over the intrinsic 
worldsheet metric was represented as a summation over two-dimensional triangulations \cite{DT,5,6,7}. These triangulations were constructed 
by gluing together equilateral triangles, which were considered flat in the interior, 
such that they formed a piecewise linear manifold with a fixed topology. The side-length
$\ep$ of the triangles served as the UV cut-off. In this way, each piecewise linear manifold (i.e., triangulation) was equipped with a geometry, different triangulations corresponded to different geometries, and one had a coordinate-free representation of the 
geometries. While the bosonic strings in physical target dimensions (i.e., dimensions larger than two) presumably do not exist
because of the existence of tachyons in the spectrum, the so-called non-critical strings, which in the Polyakov formulation 
are represented as two-dimensional quantum gravity coupled to conformal field theories with central charge $c < 1$, are well defined 
theories, 
{many aspects of which} can be found analytically, both working in the continuum \cite{KPZ,Dav88,DK89,fateev} 
and (surprisingly) also in the lattice DT formalism \cite{2dDT,more2dDT}. 
The results agree in the $\ep \to 0$ limit, thus providing explicit examples of  
successful lattice regularizations of diffeomorphism-invariant theories. A nice feature of the DT formalism is that it is 
coordinate-independent, and this includes the cut-off $\ep$. The generalization of DT to quantum gravity with Euclidean signature
in higher dimensions is in principle straightforward \cite{higherDT,Agishtein:1991cv}. {Extrapolating from the two-dimensional theory, we might thus have a four-dimensional lattice theory of gravity where the continuum limit provides us with a diffeomorphism-invariant theory of quantum gravity.} However, it has not (yet) 
been possible to find an interesting $\ep \to 0$ limit for the lattice theory \cite{firstorder,Catterall:1997xj,firstorder1,Coumbe:2014nea}. 
This led to CDT~\cite{originalCDT,Ambjorn:2005qt,ajl1,agjl,Ambjorn:2007jv}, where the starting point is a globally hyperbolic manifold and a corresponding time foliation; thus, the functional integral is then performed over Lorentzian geometries. CDT provides a lattice description of this, which involves a lattice choice of a 
proper time coordinate, but where the spatial geometries at each lattice time are represented as three-dimensional Euclidean-signature DT triangulations. While DT represents a completely coordinate-free lattice formulation of the path integral of quantum gravity, the precise continuum interpretation of CDT is still open for interpretation. It is a lattice theory which  
can be viewed as  coordinate-free in the spatial directions, while a gauge fixing
of the lab-function has been made in  the time direction. Presently it is not known if the resulting continuum theory (assuming 
it can be defined) should be viewed as a gauge fixed version of General Relativity (GR) or Ho\v{r}ava-Lifshitz gravity (HLG) 
\cite{horava1,horava2}, the later being defined precisely as a theory with a time-foliation and being invariant under spatial diffeomorphisms.  

Whether CDT is a lattice representation of GR or HLG, 
each lattice configuration that appears in the  CDT path integral is ``coordinate-free'' in the 
spatial directions. If one knows the three-dimensional triangulation of space, i.e., which simplices are glued together, one
can in principle reconstruct the exact piecewise linear spatial geometry. 
From the point of view of both GR and HLG, this seems to be
an ideal situation. Just by being given the connectivity of the triangulation (which simplices are glued together, etc.) the geometry
is given without the use of coordinates.  However, being configurations in a path integral, these geometries are quite fractal, since they incorporate quantum fluctuations at all scales, and to actually obtain a useful description of such geometries, it might be advantageous to have a suitable coordinate system. This is where the topology of space can play 
a role. If the topology of space is that of a three-torus ($T^3$), rather than the simpler topology of the three-sphere ($S^3$), 
we can use the periodic structure and the presence of non-contractible loops both to introduce useful coordinates and to map out the fractal structure of space in typical configurations present in the path integral. This is one of the reasons we have extended the original studies of CDT where space had the topology of $S^3$ to the situation where space has the topology of $T^3$;
this line of research has been pursued in \cite{pseudo-car,fractal-torus,cosmic-void,long-article}.

There are other motivations for extending the study of CDT with spatial topology $S^3$ to that of $T^3$. Contrary to four-dimensional DT, the lattice theory based on CDT has an interesting phase diagram. By that, we mean that when we vary the bare coupling constants of the theory, we observe second-order phase transitions. Such transitions are usually believed to be a necessity if  one attempts to remove the lattice cut-off $\ep$ by changing the bare coupling constants in such a way that physical observables stay fixed (and equal to the value they are assumed to have in the continuum theory). The reason is that second-order phase transitions are often related to the appearance of a divergent correlation length, and expressing physics in terms of this divergent correlation length makes it possible to move from a lattice to a continuum description. In the CDT theory, we have second-order phase transitions, which could be candidates for defining a continuum limit of the lattice theory. Usually, properties of such phase transitions are dictated by physics in the bulk, and one would expect them to be independent of the topology of the underlying spacetime. However, the phase transitions found in CDT are in many respects quite special and share some characteristics with so-called topological phase transitions studied in condensed matter systems. Thus, it was not ruled out that changing topology could also change the nature of the phase transitions. In the end, this was not what we observed, but changing the topology of space from $S^3$ to $T^3$  makes it  easier to  access certain parts of the phase diagram and thus allowed us to obtain a more complete picture of the CDT phase diagram, as is described below \cite{torus-phases1,towardsUV,torus-phases2}. 

One achievement of the study of CDT with  spatial topology $S^3$ was that for certain choices of the bare coupling constants, the measurements of the scale factor $a(t)$ of the universe, where $t$ is the above mentioned CDT proper time, could be described by a Hartle--Hawking minisuperspace model \cite{originalCDT,Ambjorn:2005qt,ajl1,agjl,Ambjorn:2007jv}. Even the fluctuations of the measured scale factor were well described by the minisuperspace model. Changing the spatial topology provides a non-trivial test of this picture, since the term in the 
Hartle--Hawking minisuperspace model that reflects the curvature of space is different for spherical and toroidal spatial topology. It turns out that in order to fit the computer data obtained for the toroidal topology of space, we have to make precisely such an adjustment of the minisuperspace action. We consider this consistency to be quite remarkable, and it should be emphasized that at no point is any background geometry put in by hand. Nevertheless, after integrating out all degrees of freedom except the scale factor, the resulting path integral results in an effective action similar to the Hartle--Hawking minisuperspace action, and the average value of the scale factor is the extremum of this action. Thus,  for both spherical and toroidal spatial topology backgrounds, ``geometries''  have emerged from the path integral, and these backgrounds are, as will be described below, very different \cite{impact-topology,kevin}. Here the  toroidal case is especially interesting, since it allows us to measure terms in the effective minisuperspace action, which are not just terms similar to the usual terms one obtains from the standard Einstein--Hilbert action by assuming isotropy and homogeneity of space. They thus represent genuine quantum effects. As we will explain below, it is not possible to measure these terms when the spatial topology is $S^3$. 

Finally, spatial toroidal topology allows us to study the effect of adding dynamical scalar fields with non-trivial winding numbers to the action, and whereas scalar fields did not have a large impact on the spacetime dynamics of CDT with spherical topology, the situation is very different when the topology is toroidal and boundary conditions of the scalar fields are non-trivial.  
  
The rest of the article is organized as follows: in Section~\ref{sec2}, we provide for completeness a short definition of CDT and discuss the CDT phase diagram as seen when the spatial topology is $T^3$. In Section~\ref{sec3}, we show that the minisuperspace description is still valid when the spatial topology is $T^3$, while  Section~\ref{sec4} discusses why the minisuperspace description, albeit quite precise, does not really reveal how spatial baby universes can dominate quantum fluctuations. Section~\ref{sec5} shows that, in the case of toroidal  topology, one can use scalar fields with non-trivial boundary conditions as coordinates, and that the use of these coordinates provides us with a detailed map of the quantum geometry. In Section~\ref{sec6}, we study the effects of including dynamical scalar fields with non-trivial boundary conditions in the action. The results seem to be quite dramatic, as the inclusion of dynamical scalar fields results in a new kind of phase transition. Finally, Section~\ref{sec7} contains a discussion.

\section{CDT and Its Phase Diagram}\label{sec2}

\subsection{Defining CDT}

\begin{figure}[t]
\centering
\includegraphics[width=0.45\linewidth]{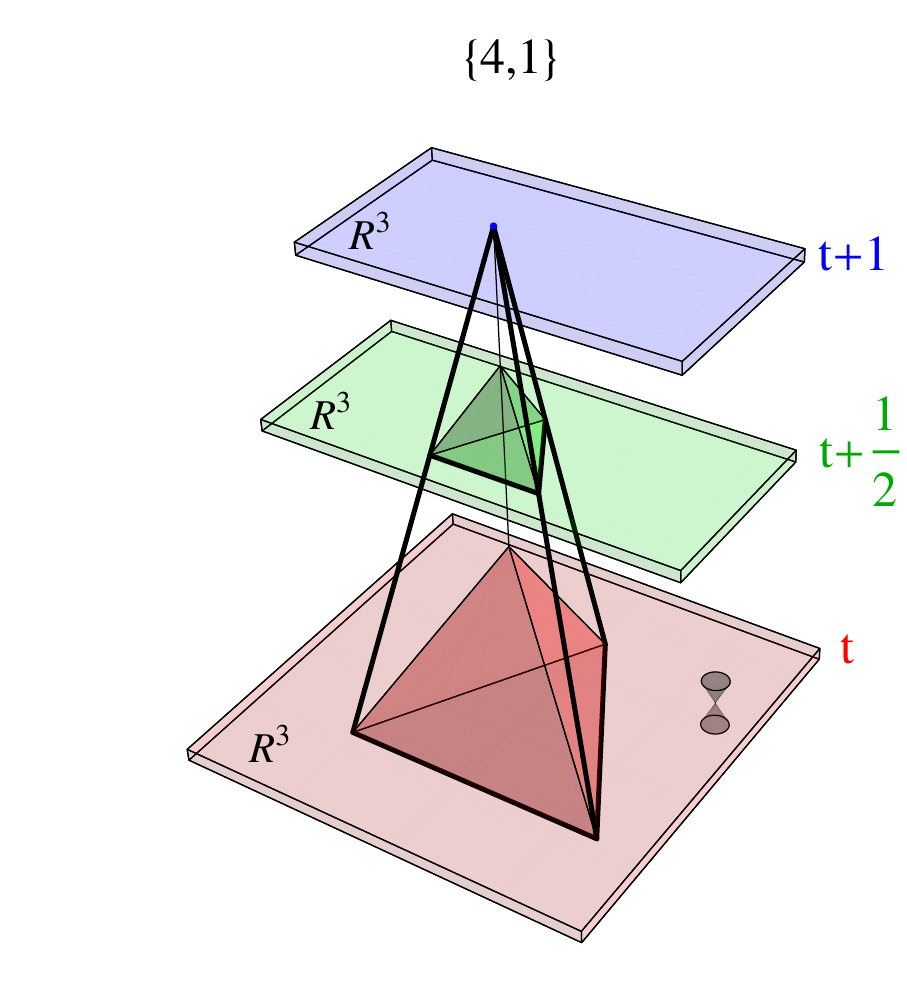}
\includegraphics[width=0.45\linewidth]{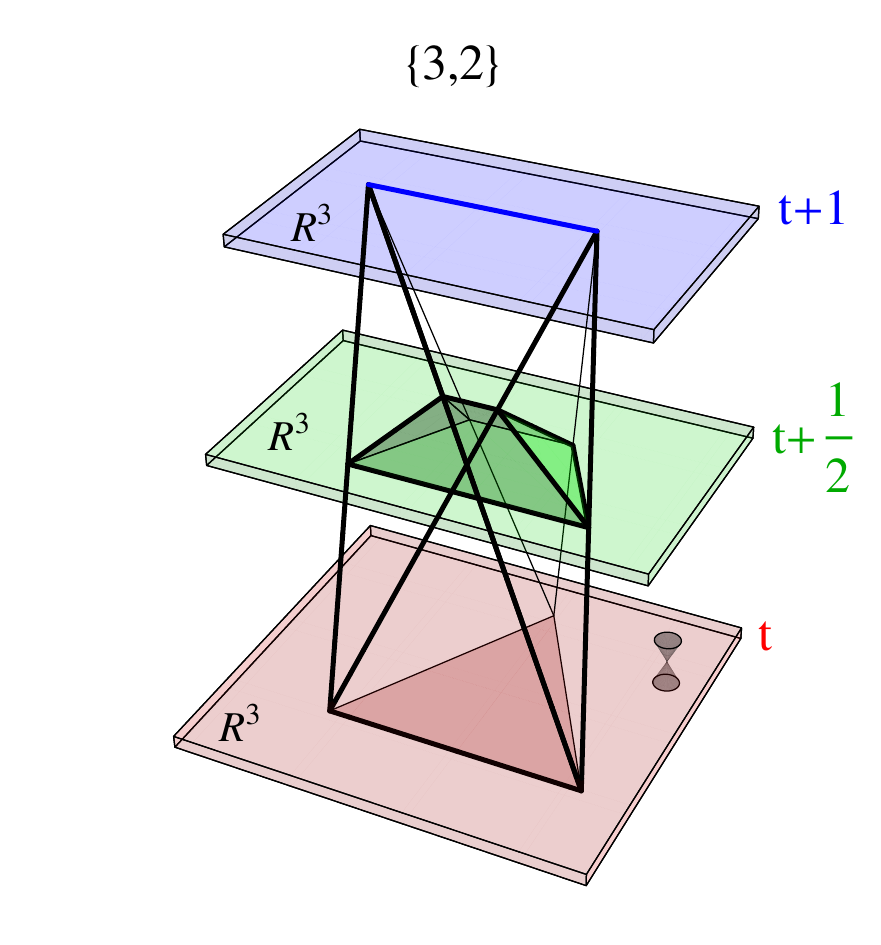}
\caption{ {Visualization of the elementary building blocks of four-dimensional Causal Dynamical Triangulations (CDT), the $(4,1)$-simplex (\textbf{left}) and the $(3,2)$-simplex (\textbf{right}). }}
\label{F:Simplices}
\end{figure}

Let us briefly describe how CDT implements the lattice version of a four-dimensional hyperbolic manifold with a Lorentzian geometry (for details, we refer to \cite{physrep}). 
First, one chooses integer time labels $t_n$. For each such time $t_n$, one constructs a three-dimensional Euclidean piecewise linear manifold  $\Sigma(t_n)$   by gluing together three-dimensional tetrahedra with link lengths $a_s\!=\! \ep$. All   $\Sigma(t_n)$  have the same topology, which we denote $\Sigma_{top}$. The four-dimensional slab between two  three-dimensional triangulations at $t_n$ and $t_{n+1}$ is now ``filled out'' by four-simplices that have four vertices at $t_{n}$ (i.e., they form one of the tetrahedra there) and one vertex at $t_{n+1}$. We denote such a four-simplex a (4,1)-simplex. Alternatively, a four-simplex can have three vertices at $t_{n}$ (which form a triangle, shared by two of the tetrahedra  at $t_n$) and two vertices at $t_{n+1}$ (which form a link shared by a number of tetrahedra at $t_{n+1}$). We denote such a four-simplex a (3,2)-simplex. The construction is illustrated in Figure~\ref{F:Simplices}. Analogously, one can construct (2,3)-simplices and (1,4)-simplices. For each of these four-simplices, the links connecting $t_n$ and $t_{n+1}$ are considered time-like with proper length squared $a^2_t = -\a\, \ep^2$, $\a > 0$. The gluing of these four-simplices should be such that the topology of the slab between $t_n$ and $t_{n+1}$ is $\Sigma_{top} \times [0,1]$. In the CDT path integral, one now performs a summation over all possible three-dimensional triangulations for each of the 
time slices with topology $\Sigma_{top}$ and all possible ways to connect two neighboring time slices in the way described above. The lattice spacing $\ep$ is the UV cut-off in the path integral. In a continuum path integral, we would like to use  the Einstein--Hilbert action. There might be good reasons to also include higher derivative terms, but here we will restrict ourselves to the Einstein--Hilbert action, which has a natural geometric representation on piecewise linear manifolds, known as the Regge action \cite{regge} (for attempts to add in an explicit way higher curvature terms to the Regge action,  see~\cite{highercurvature}). One important aspect of the CDT construction is that we, for each four-dimensional triangulation, can perform a rotation to Euclidean geometry by rotating $\a \to -\a$ via the lower complex plane. By such a rotation, we have the following transformation of the Regge action for a Lorentzian CDT triangulation $T_L$ to an Euclidean triangulation  $T_E\;$: $i S_R(T_L) \to -S_R(T_E)$. Because our piecewise linear manifold is constructed by gluing together only the two kinds of four-simplices shown in Figure~\ref{F:Simplices}, the Regge action becomes very simple, and remarkably, it depends only on the total number of (4,1)-, (1,4)-, (3,2)- and (2,3)-simplices, the number of vertices $N_0$, the asymmetry parameter $\a$ and the bare lattice coupling constants. After the rotation to an Euclidean triangulation $T_E$, the Regge action can be written as   
\beql{SRegge}
S_{R}[T_E]=- (K_{0} + 6 \Delta) N_{0}+K_{4}\left(N_{(4,1)}+N_{(3,2)}\right)+\Delta \ N_{(4,1)}.
\eeq   
where $N_{(4,1)}$ is the number of (4,1)- and (1,4)-simplices and similarly $N_{(3,2)}$  denotes the number of (3,2)- and (2,3)-simplices. $\Del$ is a rather complicated function of $\a$ and it is called the asymmetry parameter. $\Delta =0$ corresponds to $a_s = a_t$. Finally, $K_0$ is proportional to the inverse of the bare, dimensionless lattice version of Newton's gravitational coupling constant, and $K_4$ can be related to the bare lattice cosmological constant.  While $\Del$ appears as an asymmetry between spatial and time directions that we have put in by hand, we will treat it as an additional coupling constant. Thus, our model has three coupling constants, $\Del$, $K_0$ and $K_4$.

After the rotation from Lorentzian triangulations $T_L$ to Euclidean triangulations $T_E$, we can view the path integral as a partition function for a statistical theory of random geometries, but note that the set of Euclidean triangulations we obtain this way is different from the set of Euclidean triangulations one constructs in DT, since they have a time foliation.  We now write:
\beql{ZE}
{ Z}_{CDT} = \sum_{ T_L}\e^{i S_{R}[ T_L]}  \xrightarrow[\alpha \, \to \, -\alpha]{} { Z}_{CDT} = 
\sum_{ T_E}\e^{-S_{R}[{ T_E} ]} \ .
\eeq
The advantage of this rotation is that we can study the partition function using Monte Carlo simulations, which we do in the rest of this article. Such Monte Carlo simulations can be used to generate a sequence of spacetime configurations, where the relative probability of a given configuration (i.e., an Euclidean CDT triangulation $T$) will be given by the Boltzmann factor $e^{-S_R[T]}$. Thus, one can use a sequence of such configurations to calculate expectation values of spacetime ``observables'', or one can try to study specific features of the configurations.

\subsection{The CDT Phase Diagram}

\begin{figure}[t]
\includegraphics[width=0.95\linewidth]{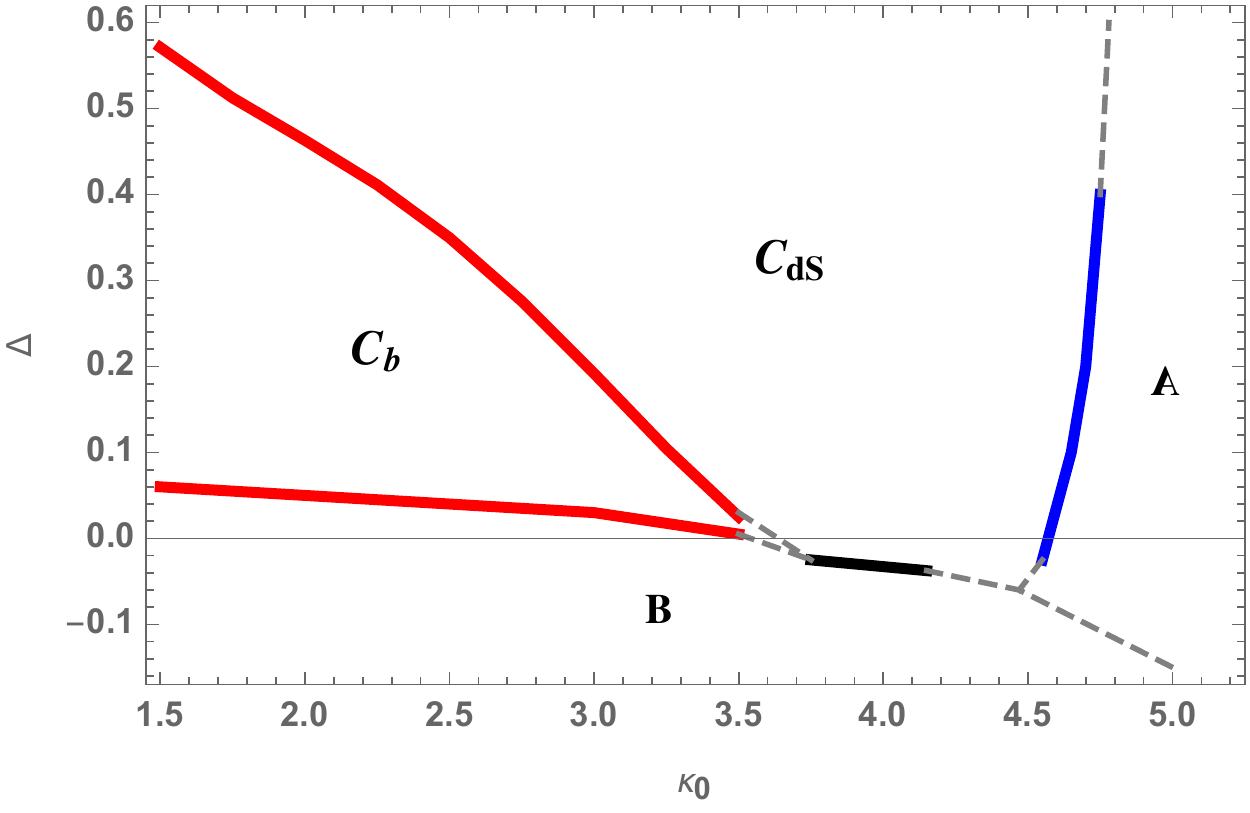}
\caption{The CDT phase diagram. The phase transition between phase $C_{dS}$ and $C_b$ is (most likely) second order, as is the transition between $C_b$ and $B$, while the transition between $C_{dS}$ and $A$ is first-order. The transition between $C_{dS}$ and $B$ is still under investigation, and so is the $A-B$ phase transition.}
\label{Phasediagram}
\end{figure}

As mentioned above, one of the first tasks in a lattice field theory is to determine the phase diagram, i.e., to understand where there can be phase transitions as a function of the bare coupling constants of the theory. Values of the bare coupling constants where a second order phase transition occurs are points where one might attempt to define a continuum limit of the lattice field theory. We study the theory using Monte Carlo simulations. Thus, we are always bound to have a finite spacetime volume. There are two implications of this. First of all, real phase transitions only take place at infinite spacetime volume, so what we observe 
for finite volume is a ``pseudo-transition''. Sometimes it is not easy to decide if the system really moves towards a critical behavior of the variable chosen to characterize the transition when the size of the system goes to infinity, or whether one is just observing some rapid ``cross-over'' of that variable. It might require a systematic study of the system as a function of the lattice size $N$. For a theory of gravity, there is the additional twist that the volume of spacetime itself is a dynamical variable. In the Monte Carlo simulations, the volume $N_4$, the number of four-simplices forming the spacetime, fluctuates. In the Monte Carlo simulations, we have often found it convenient to fix $N_4$ or $N_{(4,1)}$  (to the extent 
{allowed by}
the computer algorithms 
{we use to evolve the triangulations).}
We compensate for this by performing computer simulations for a whole range of $N_4$. From the action \rf{SRegge}, it is seen that keeping $N_4$ or $N_{(4,1)}$ fixed implies that $K_4$ is not playing a role in the phase diagram, which is thus two-dimensional, depending on $K_0$  and $\Del$. Despite the very simple nature of the action \rf{SRegge}, the phase diagram is quite complicated, as shown in Figure~\ref{Phasediagram}. The red curves show what is believed to be second-order phase transitions, the blue curve shows a first order transition, and the black curve a transition 
{of either first or second order, to be determined in further studies.}
Most of the phase diagram was 
{established} some time ago, using the original CDT set-up, where the constant time slices $\Sigma(t)$ had the topology of $S^3$. However, the neighborhood of the black curve, which is perhaps the most interesting part, was only reachable in the Monte Carlo simulations after we changed the topology of $\Sigma(t)$ to $T^3$. We will discuss below why there could be such a dependence on topology.

The phase transitions in lattice gravity are unusual, since they are transitions involving spacetime itself. Usually, phase transitions of a statistical system take place in a fixed spacetime background, and higher-order transitions are typically defined by a divergent correlation length associated with an order parameter (e.g., the correlation length of spin correlation functions for ferromagnetic systems). Measurements also take place on a fixed spacetime background, which of course is most often just flat spacetime. In lattice gravity, however, it is not clear what order parameter to use and whether or not it makes sense to talk about a correlation length of a wildly fluctuating geometry. In view of this, it is maybe unsurprising that the phase transitions we have observed have been somewhat unusual. Nevertheless, it has to some extent been possible to apply the usual understanding of phase transitions, as we will now explain. 

Our present understanding of the phases is the following. What is denoted as the $C_{dS}$-phase (the so-called de Sitter phase; sometimes denoted as just the C phase)  in Figure~\ref{Phasediagram} is characterized by the homogeneity and isotropy of the spatial triangulations of $\Sigma(t)$ in the limit where the number of tetrahedra goes to infinity. The phase transition between  phase  $C_{dS}$ and the so-called bifurcation phase $C_b$ breaks this symmetry. Special vertices of high order (i.e., vertices sharing a large number of simplices) start to develop, and they are {only} loosely connected over time $t$. As we move towards the phase denoted $B$, the order of these singular vertices grows, and the universe starts to contract in the time direction. The phase transition between the $C_b$ and the $B$ phase is one where the time dimension effectively disappears, with most of 
the $N_{(4,1)}$ simplices incident on  a single time-slice, i.e., a phase transition of dimensional reduction. Finally, the phase transition between phase $C_{dS}$ and the so-called phase $A$ also has an interpretation as a change of geometry. Phase $C_{dS}$ is characterized by a strong correlation between neighboring time slices, and it is this correlation that ensures that the geometries in the $C_{dS}$ phase have an interpretation as a genuine four-dimensional geometry (as we will discuss in detail in the next Section). However, the transition to phase $A$ is one where this correlation is lost, and phase $A$ can, loosely speaking, be viewed as a time-sequence of disconnected spatial universes, and one interpretation is that, moving into phase A, the conformal factor instability of Euclidean gravity
will dominate.
The transition between phases $A$ and $B$ is probably not so interesting from
the point of view of physics, but the point where the C, B and A phases meet {\it is} very interesting, as it can be  a potential fixed point of the theory and can thus  play an important role in understanding how to obtain the continuum limit of the lattice theory.

 
\begin{figure}[th]
\includegraphics[width=0.32\linewidth]{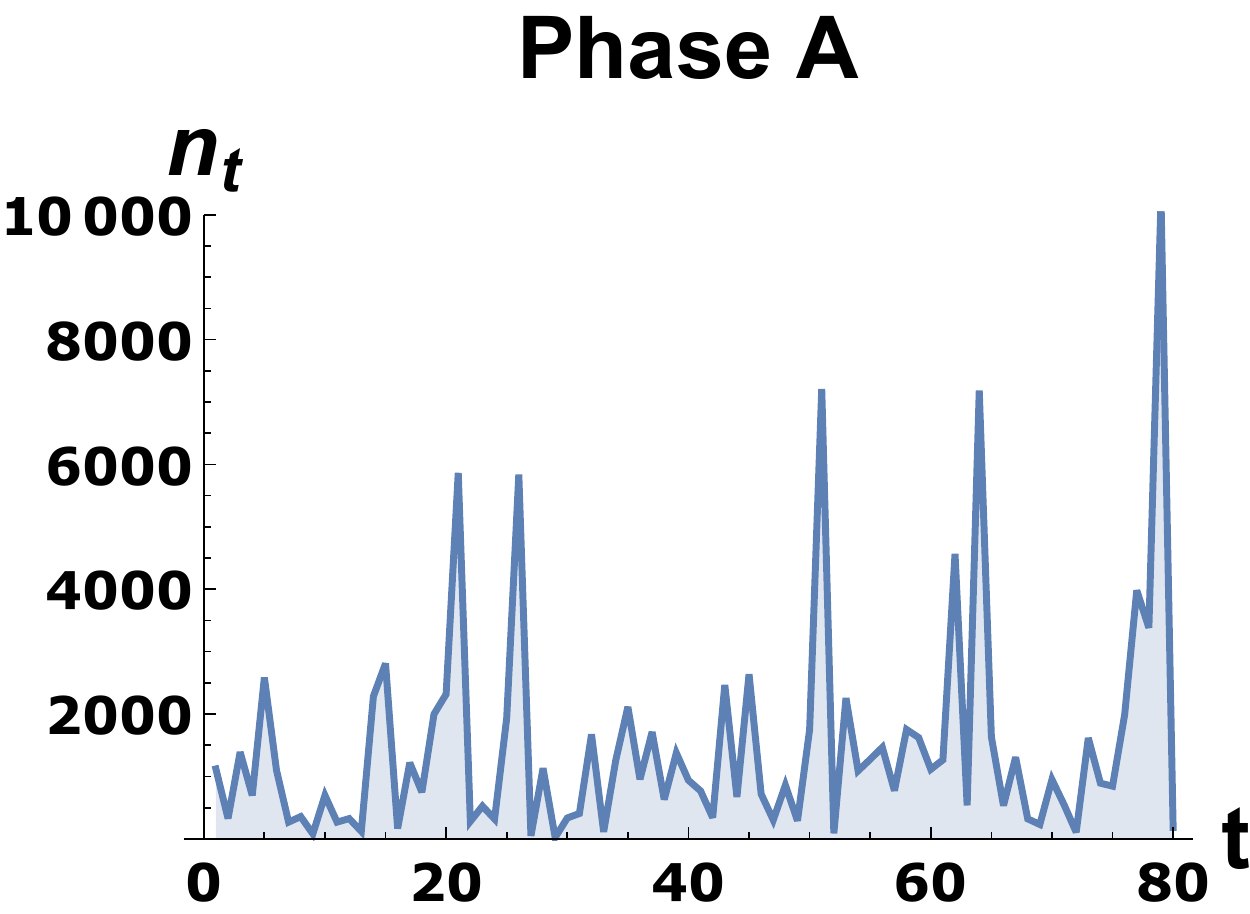}
\includegraphics[width=0.32\linewidth]{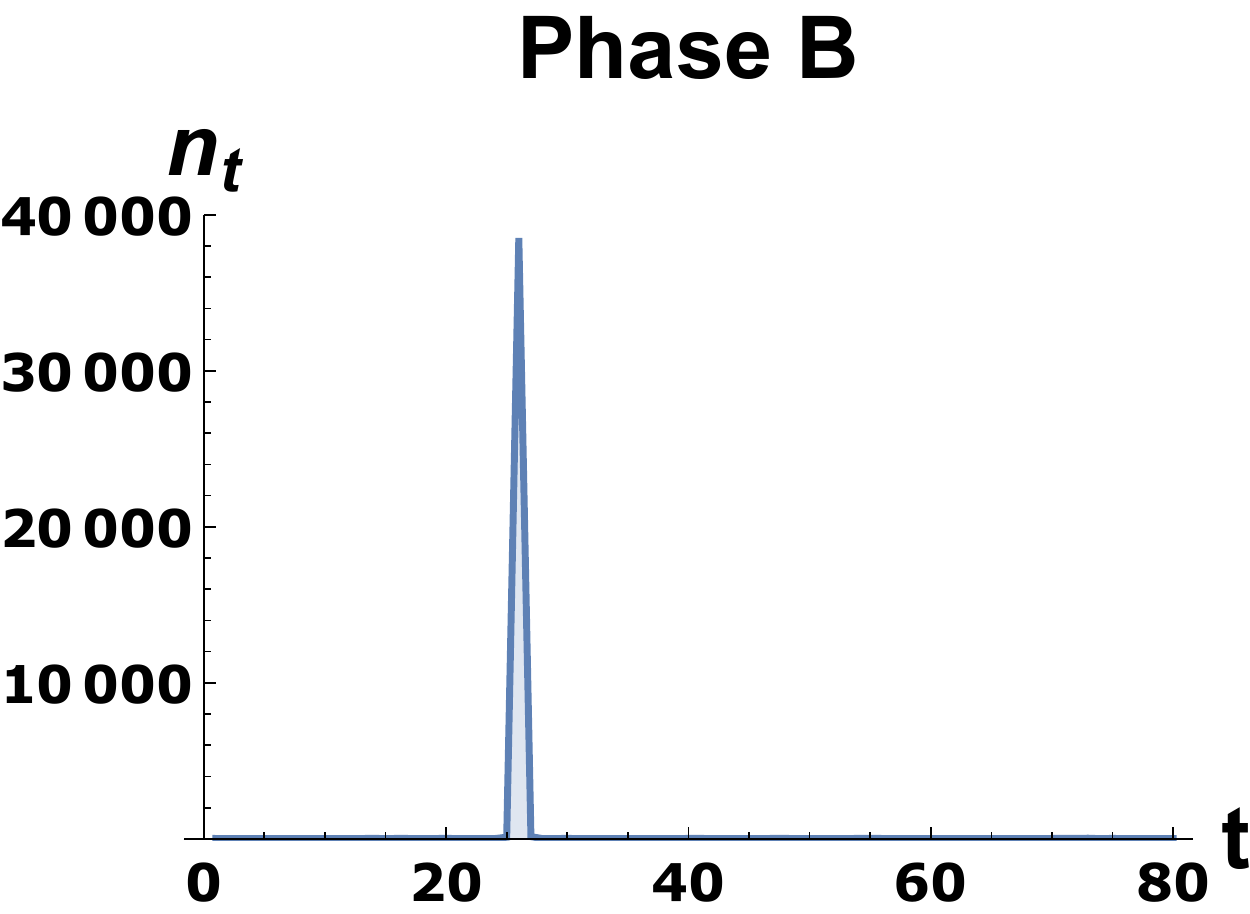}
\includegraphics[width=0.32\linewidth]{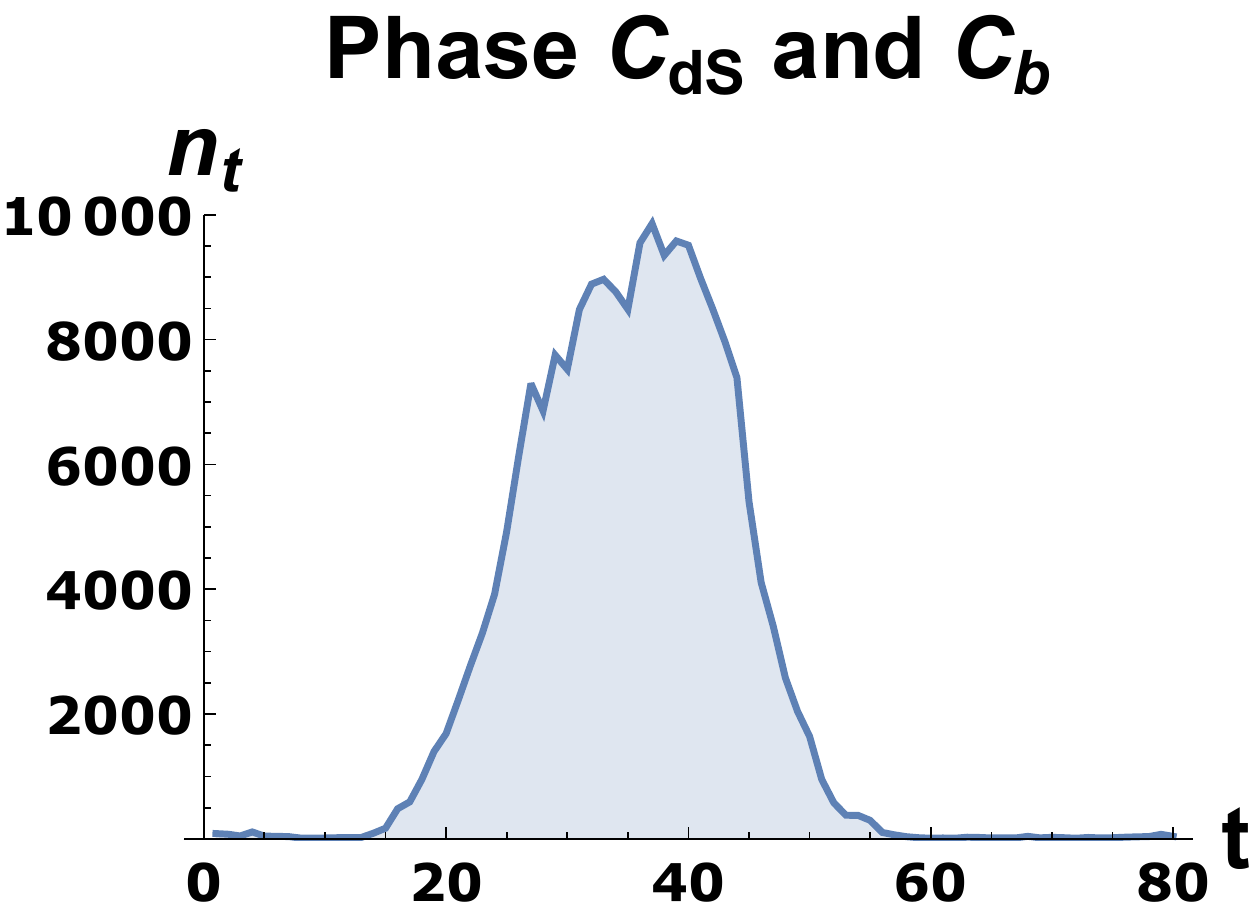}
\caption{{{Typical volume profiles} $n_t=N_{(4,1)}(t)$ in various CDT phases in spherical spatial topology $\Sigma=S^3$.}}
\label{F:VolSphere} 
\end{figure} 
\vspace{-6pt}
 \begin{figure}[h]
\includegraphics[width=0.3\linewidth]{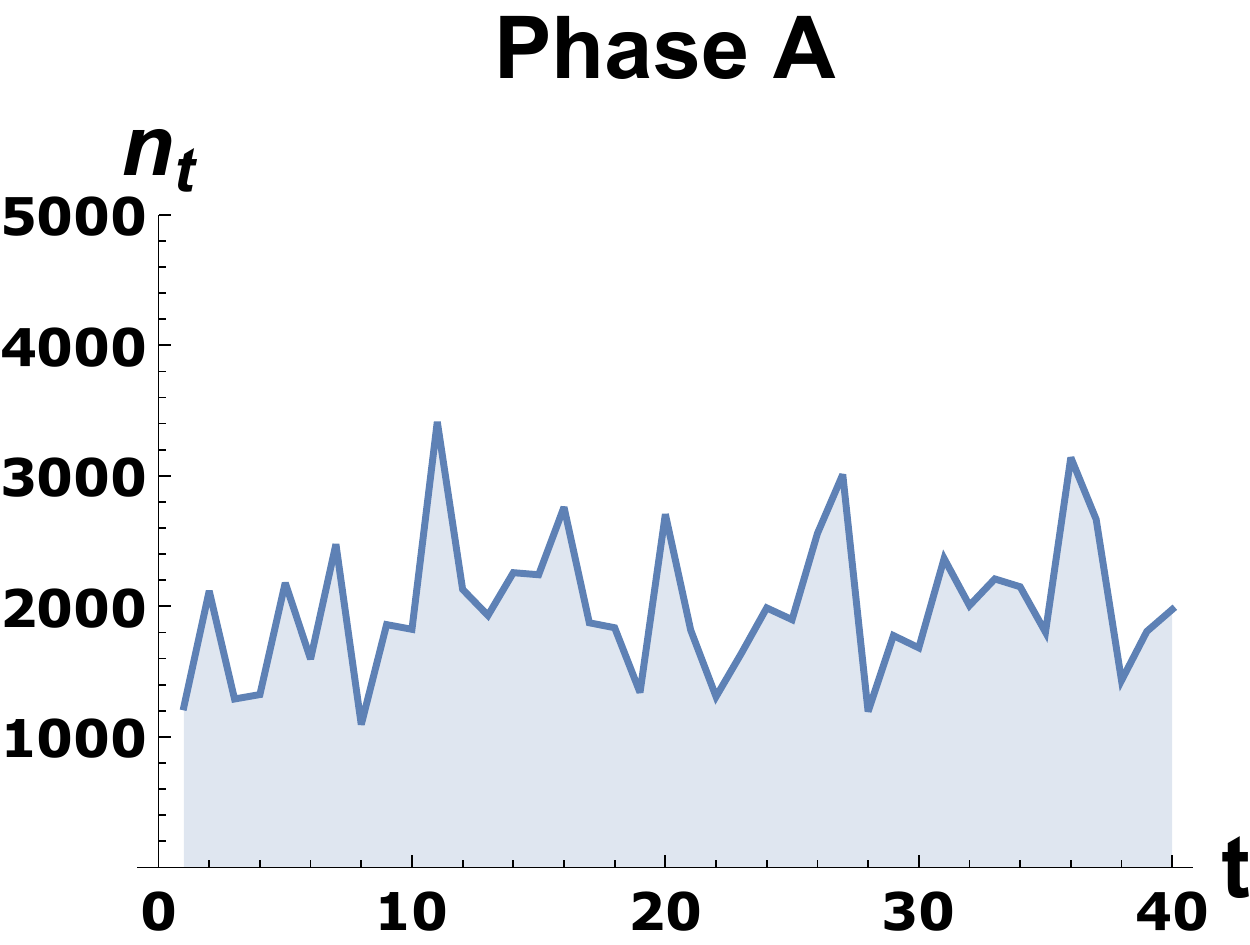}
\includegraphics[width=0.3\linewidth]{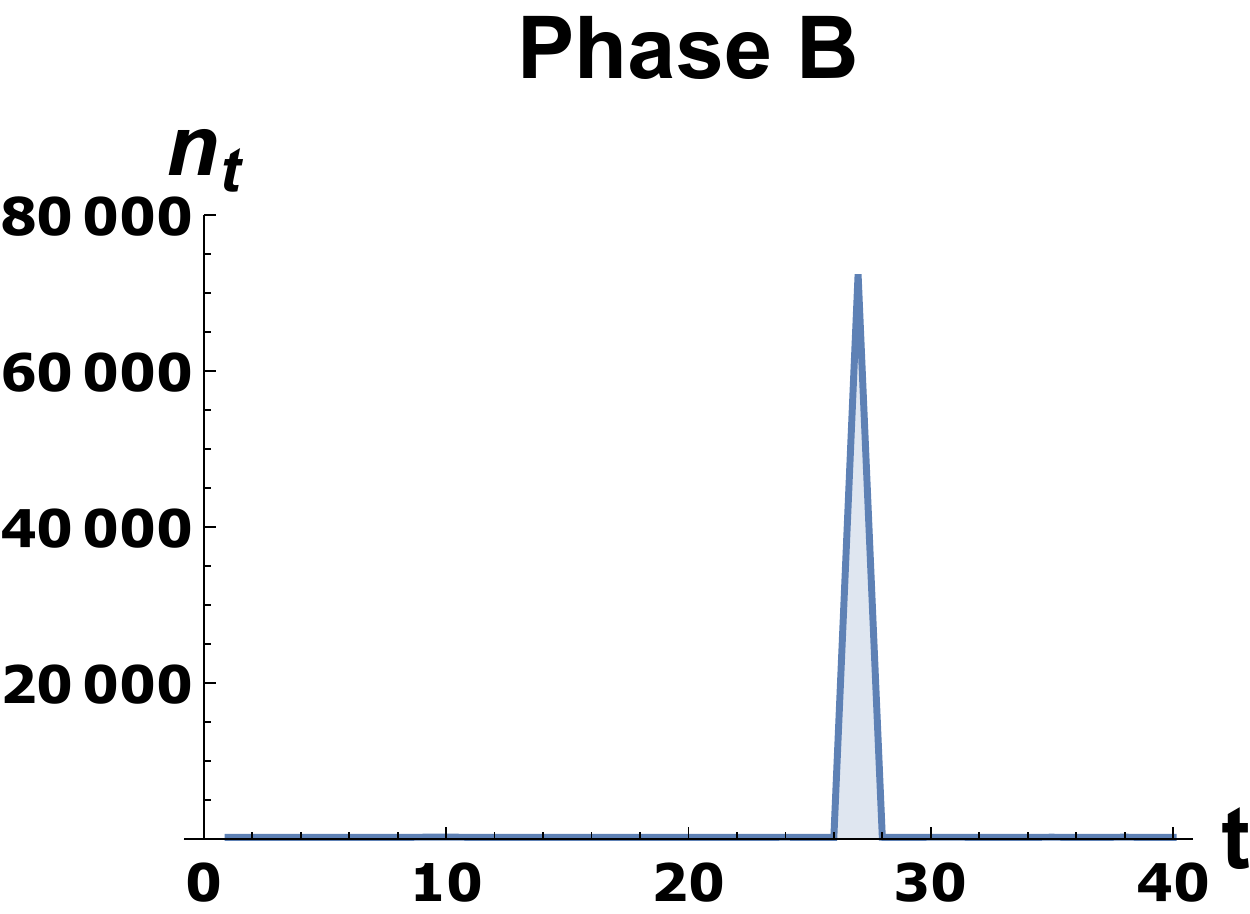}
\includegraphics[width=0.3\linewidth]{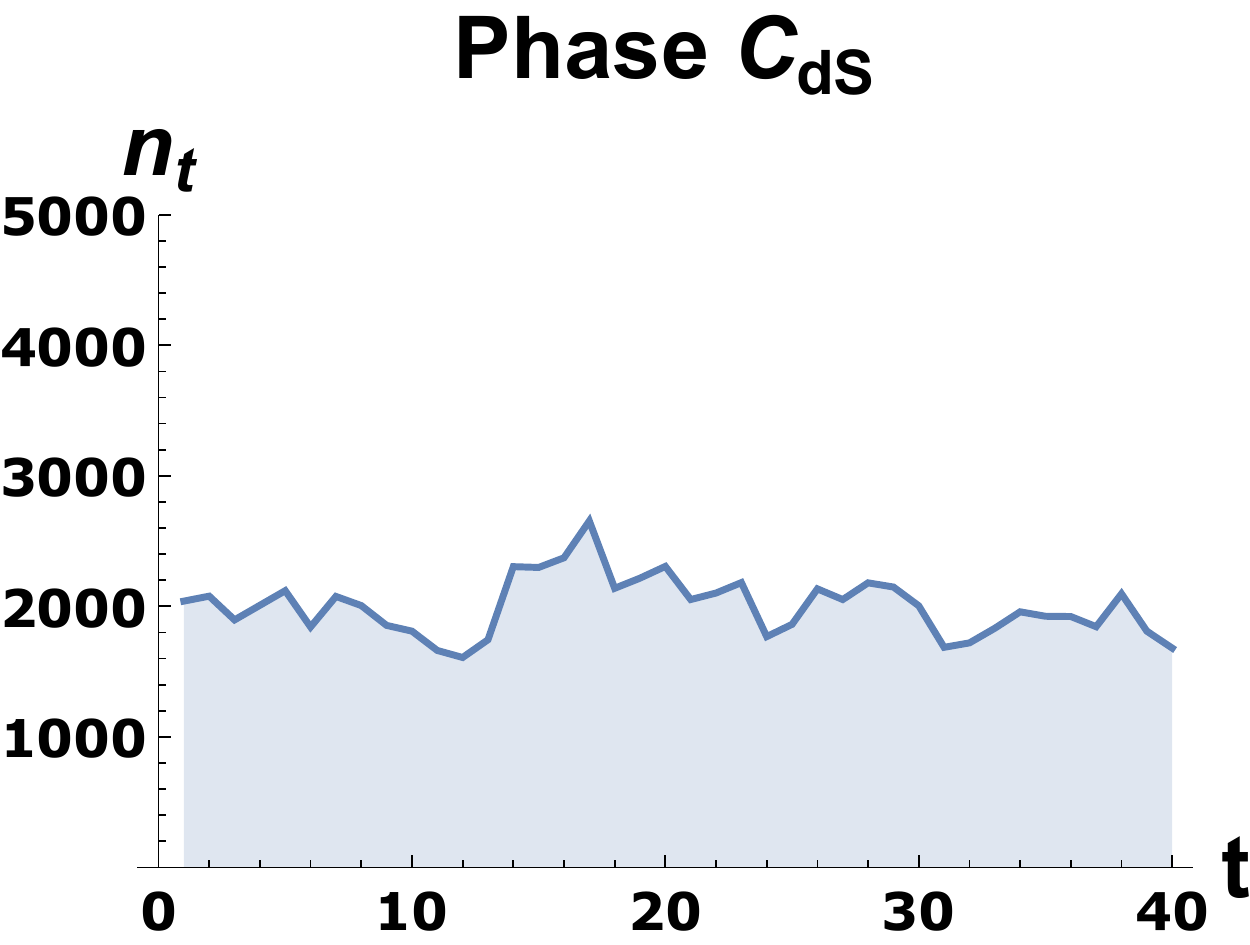}
\caption{ {{Typical volume profiles} $n_t= N_{(4,1)}(t)$ in  various CDT phases with toroidal spatial topology $\Sigma=T^3$. The volume profile in the $C_b$ phase is similar to the profile 
for  $\Sigma =S^3$}.}
\label{F:VolTorus}
\end{figure}

\begin{figure}[t]
\centering
\includegraphics[width=0.7\linewidth]{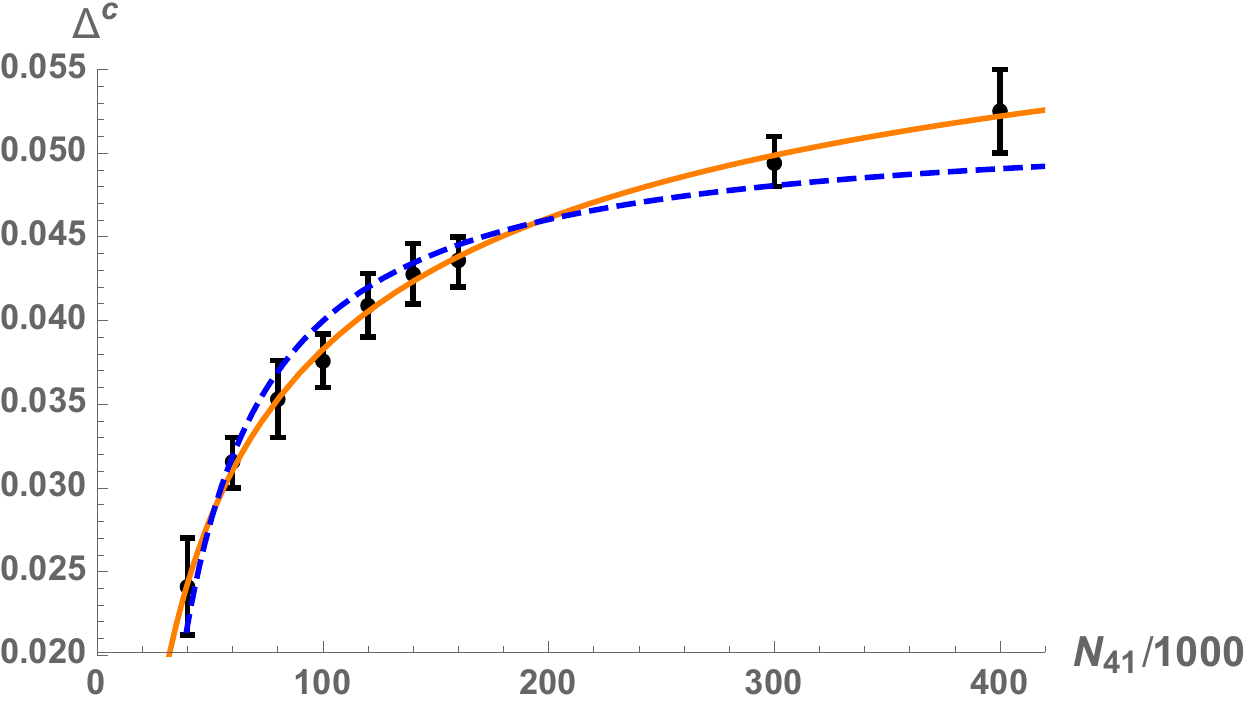}
\caption{Lattice volume dependence of the pseudo-critical $\Delta^c(N_{(4,1)})$ values in CDT with toroidal spatial topology and for fixed $K_0=2.2$ together with the fit of the finite size scaling relation \rf{ja1} with critical exponent $\nu=2.7$ (orange solid line) and the same fit with a forced value of  $\nu=1$ (blue dashed~line). }
\label{Figfit}
\end{figure}

{The different types of spatial volume profiles  just discussed are shown in \mbox{Figures~\ref{F:VolSphere} and \ref{F:VolTorus}}.} They show  $N_{(4,1)}(t)$ for typical triangulations generated by  Monte Carlo simulations in the various phases. In phase $C_{dS}$, it is seen that the universe has a well-defined extension both in space and time directions. This is true both for the spatial topology $S^3$ and $T^3$, although the average profiles around which the spatial volume fluctuates are very different in the two cases. {{In the case of }$\Sigma=S^3$ the spatial volume profile in phase $C_b$  is seemingly very similar to the profile in phase $C_{dS}$. However, a closer analysis shows that the time extension of a ``blob'', like the one shown in Figure~\ref{F:VolSphere} (right), in phase $C_b$ does not scale in the correct way in the time direction when one increases the size $N_4$ of the triangulation. If the spatial topology is $\Sigma=T^3$, the volume profile in phase $C_b$ has a ``blob'', like for topology $S^3$, and thus looks increasingly different from the volume profile in phase $C_{dS}$ when one approaches phase $B$}. We will discuss this in the next section. Although the phase transitions observed in CDT have a relatively transparent physical interpretation as being related to a change in the typical geometry of configurations dominating the partition function when one changes the coupling constants, i.e., a competition between entropy of a certain kind of configurations and weight they receive via their action, the characteristics of the phase transitions are somewhat atypical. One atypical feature is that the second-order transitions for smaller systems show hysteresis and atypical double peak structures {for histograms} of some of the observables. These are  features that are associated with first-order transitions, and they usually become more pronounced when the system size increases. Here, the opposite often happens: the double peaks and the hysteresis disappear with increasing system size. Furthermore, when we analyze the change in the position of the pseudo-critical points as a function of the system size, the critical exponents associated with such a change favor higher order transitions for the transition lines shown in red in Figure~\ref{Phasediagram}. As an example, one can choose as an order parameter the ratio $OP_2 = N_{(3,2)}/N_{(4,1)}$ and study the change of $\la OP_2(K_0,\Del)\ra$, as well as the so-called susceptibility $\chi_{OP_2} = \la OP_2^2  \ra - \la OP_2\ra^2$, {when $K_0$ and/or $\Del$ are changed}. One usually identifies the pseudo-critical points as the values $K_0, \Del$ where $\chi_{OP}$ has a peak (assuming that $OP$ is a good order parameter for the transition). Figure~\ref{Figfit} shows the results of such measurements for the $C_b-B$ transition for toroidal topology for the case where $K_0$ is kept fixed and $\Del$ is varied until one finds the maximum of $\chi_{OP_2}$, thus identifying the pseudo-critical 
 $\Del^c(N_{(4,1)})$. 
 
  The fit is of the form 
 \beql{ja1}
 \Del^c(N_{(4,1)}) = \Del^c (\infty) - {C}\,{N_{(4,1)}^{-1/\nu}},
 \eeq
 {where $\nu =1$ is the exponent for first-order transitions and $\nu > 1$ is expected for higher-order transitions.}

 To understand  why some of the results of  the numerical simulations  are ambiguous, it is important to understand the nature of the simulations: they are Monte Carlo simulations where one attempts to locally change the triangulations, and where the change is accepted or rejected according to the so-called Metropolis algorithm {with a probability} depending on the action of the configurations. Eventually, starting with one triangulation and making a number of such local changes, one will generate another statistically independent triangulation, with the relative probability dictated by the exponential of the action. However, even with a local updating algorithm that is ergodic, i.e., whose repeated application allows one to reach any configuration {with the same topology}, many such local updates might be needed. This problem is well known near phase transitions
 {in simpler statistical systems, and it might occur here as well.}
 Our present understanding is that the creation of 
 {high-order vertices} might be easy using a local updating algorithm, but resolving them might take many attempts, as it might require a major rearrangement of the triangulation. This might explain why certain rearrangements might be easier 
 {in the case of larger triangulations, for which the hysteresis disappears}.
 It might also explain why certain aspects of the phase transitions look different {depending on} {whether} we impose spatial topology $S^3$ or $T^3$. As already mentioned, we are only able to reach the region near the $C_{dS}$ and $B$ transition using toroidal topology. 
 In the case of spherical topology, the computer time it takes to generate genuinely independent triangulations becomes too long for 
 large triangulations.

\section{The Effective Minisuperspace Action}\label{sec3}

Let us consider the computer simulations of 
a quantum universe, for $N_4$ (or for convenience, often $N_{(4,1)}$) fixed, for a choice  of coupling constants $K_0, \Del$ such that we are in the de Sitter phase, and with periodicity in the time direction assumed. In the case where the spatial topology is that of $S^3$, this 
{assumption is made} mainly for the convenience of the computer simulations, and it plays no role in the numerical results if the total length $T$ of the time {coordinate} is large enough. In the case of toroidal spatial topology, however, it is {not only convenient, but also} natural. From plots of the spatial volumes $n_t \!=\! N_{(4,1)}(t)$ as functions of $t$, as shown in  Figures~\ref{F:VolSphere} and \ref{F:VolTorus}, it is clear that the volume profiles $n_t$  
are very different when space has the topology of 
{$T^3$ rather than that of $S^3$.}
The profiles are also relatively well defined; i.e., it makes sense to attempt an interpretation of these profiles as a ''background profile'' with superimposed quantum fluctuations. One can try to substantiate such a picture by a careful study of the $\la n_t\ra $ and $\la n_t n_{t'}\ra$. In this way, we are basically studying the quantum mechanics of a single variable, namely the ``scale factor'' $a(t) := n^{1/3}_t$ of the universe. Note that  we are not imposing a minisuperspace restriction on our theory, but we are simply integrating out all degrees of freedom other than $a(t)$ in the path integral. A priori, it is not clear that one will obtain any useful description of the system in terms of $a(t)$, or that it makes sense to talk about a  ``background profile'' for $a(t)$, since no background geometry is put into the path integral.

\begin{figure}[t]
\includegraphics[width=0.5\textwidth]{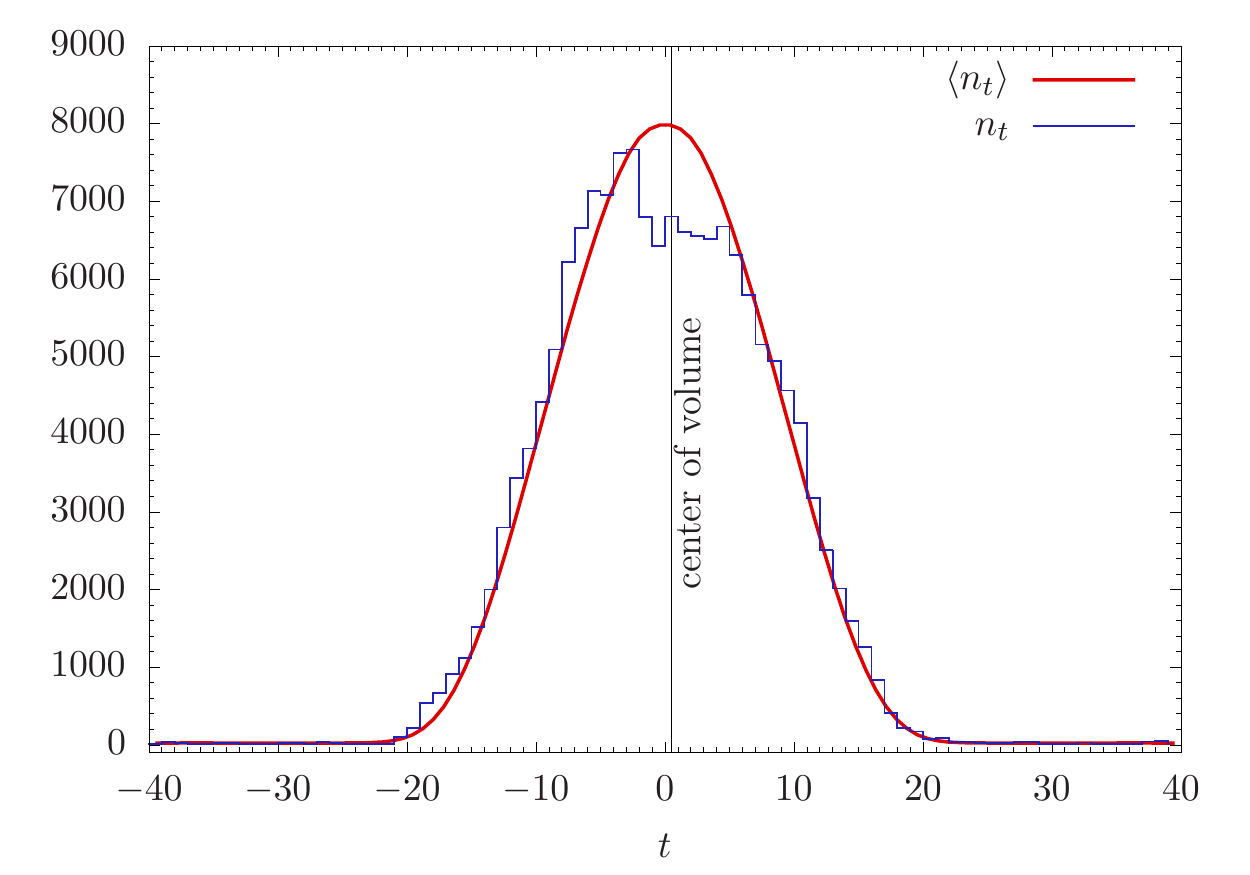}
\includegraphics[width=0.5\textwidth]{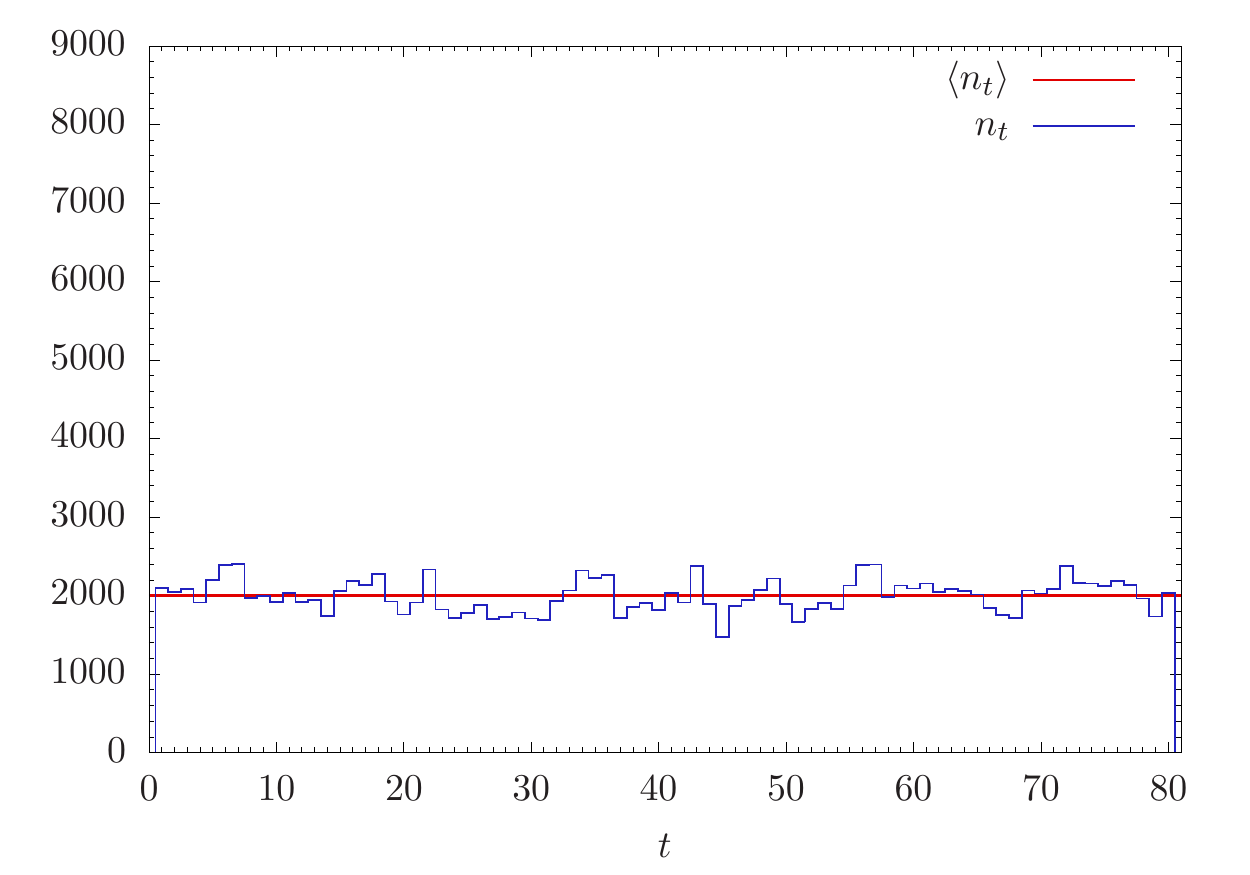}
\caption{The average (red) and typical (blue) volume profiles for spherical spatial topology
(left figure) and for toroidal  spatial topology (right figure) in the $C_{dS}$ phase. 
In both cases, the time direction has length T = 80 (in lattice units) with periodic boundary conditions. }
\label{fig:spheretorus}
\end{figure}

$\la n_t \ra$ is shown in Figure~\ref{fig:spheretorus}, together with profiles
for typical configurations, for $\Sigma$ with the topology of $S^3$ and $T^3$, so that one can assess the size of the fluctuations 
around $\la n_t \ra$ for the particular size of triangulations used.

In the case of the sphere, the part of the volume profile where $\la n_t \ra$ is significantly larger than the minimal value (i.e., five tetrahedra forming a triangulation of $S^3$) is given by the formula 
\beql{ja3}
\la n_t \ra \propto  N_{(4,1)}^{3/4}  \cos^3 \left( c \;\frac{t}{ N^{1/4}_{(4,1)}} \right).
\eeq
The constant $c$ in the above expression is independent of $T$ and $N_{(4,1)}$. 
The important point is that {the} time-extent of the ``blob'' is independent of $T$ (for sufficiently large $T$), and thus the 
time-extent of the blob will grow as $N_{(4,1)}^{1/4}$, showing 
that the blob represents a genuine four-dimensional ``sphere''.
In the case of the torus, we have on the other hand,
\beql{ja4}
\la n_t \ra \propto \frac{N_{(4,1)}}{T},
\eeq
and thus the spatial volume profile depends on the choice of $T$.

Since the fluctuations are relatively small, and increasingly so, for increasing  $N_{(4,1)}$, it makes sense to construct a minisuperspace action $S[n_t]$ such that the extremum of $S[n_t]$ is $\la n_t \ra$ and the fluctuations 
$\eta_t = n_t - \bar n_t$ are described by 
the quadratic expansion of  $S[n_t]$  around  $\bar{n}_t = \la n_t \ra$. Thus, we write
\beq\label{j2} 
{S[n = \bar n + \eta] = S[\bar n] + \frac{1}{2} \sum_{t,t'} \eta_t P_{t t'} \eta_{t'} + O(\eta^3).}
 \eeq
 To quadratic order in the fluctuations we can determine $P_{tt'}$ by measuring the covariance matrix 
 \beq\label{j1}
 C_{t t'} = \langle \eta_t \eta_{t'} \rangle = \left\langle (n_t - \langle n_t \rangle) (n_{t'} - \langle n_{t'} \rangle) \right\rangle.
 \eeq
 To quadratic order in $\eta_t$,  $P$ is the inverse of the covariance matrix:
\beq\label{j3}
 P_{t t'} = [C^{-1}]_{t t'} = \left. \frac{\partial^2 S[n_t]}{\partial n_t \partial n_{t'}}\right|_{{n_t =\bar{n}_t }}. 
 \eeq
  In this way, a measurement of the covariance matrix in principle allows us to determine the minisuperspace action $S[n_t]$ up to quadratic order in the fluctuations. In practice, there are a number of technical issues, which
 {depend somewhat on whether $\Sigma(t)$ has the topology of $S^3$ or that of $T^3$.}
 For the case of $S^3$, we refer the reader to the original articles or to the review \cite{physrep}, while details for $T^3$ can be found in \cite{impact-topology,kevin}. Let us here just summarize the results in the following formula:
 \begin{equation}\label{j20}
{S[n_t] = \sum_t\left[ \frac{1}{\Gamma} \frac{(n_{t + 1} - n_t)^2}{n_{t + 1} + n_t} + 
\alpha \,n_t^{1/3}+ \mu \, n_t^{\gamma} +\lambda \, n_t\right] .}
\end{equation}

It should be mentioned that it is possible to derive \rf{j20} by an independent method, which we refer to as the transfer matrix method. It studies the transition amplitude between neighboring time-slices and relates the amplitude to an effective action (for details, see~\cite{transfer,transfer2} for the general description and the use in the case of  $S^3$, and \cite{kevin} for the use in the case of $T^3$). Within measurement precision, the action \rf{j20} can reproduce our measurements of $\la n_t \ra$ and $\la n_t n_{t'}\ra$. 

In the effective action $S[n_t]$, the first term is a discretized version of a kinetic term, and the coefficient $\Gamma$ is the same for both spherical and toroidal spatial topology. $\Gamma$ depends on the bare coupling constants used in the simulation and can be viewed as proportional 
to a dimensionless lattice gravitational coupling constant. The next term is 
linear in the scale factor $a(t) = n_t^{1/3}$. As we will discuss shortly, its origin in a minisuperspace model of the universe, where one assumes isotropy and homogeneity, is that of intrinsic curvature of the three-dimensional space at time $t$, i.e., a term $V_3(t) R_3(t) \propto a(t)$, where $V_3(t)$ is the three-dimensional volume and $R_3(t)$ is the scalar curvature on the three-dimensional space at $t$. In the case of $S^3$, such a term is present, and we find indeed that $\a$ is not zero when we construct $S[n_t]$. For the torus, one has $R_3(t) =0$, and we find  that $\a\! = \! 0$ is consistent with our data for $T^3$.  The coupling constant $\lambda$ has a minisuperspace interpretation as proportional to a cosmological constant. However, as long as we fix the four-volume in our computer simulations, $\lambda$ does not really play the role of a physical cosmological constant, but rather of a Lagrange multiplier, which ensures that the extremum of $S[n_t]$ is the $\bar{n}_t$ we observe. In the case of $S^3$, this implies that $\lambda \propto  - 1/N_{(4,1)}^{1/2}$. In the $T^3$ case, the constant solution shown in Figure~\ref{fig:spheretorus} corresponds to $\lam\!=\! 0$ if we ignore the term $\mu\, n_t^\gamma$ in \rf{j20}. Let us now turn to this term. As we will show below, all terms on the right-hand side of \mbox{Equation~\rf{j20}} appear in a classical GR minisuperspace action, except the term $ \mu \, n_t^\gamma$. This term thus seems to be a genuine quantum term. Unfortunately, it is subleading compared to the  term $\a \, n_t^{1/3}$ in the case of the sphere. Thus, we cannot really determine it reliably from the data. However, in the toroidal case, $\a \, n_t^{1/3}$ is absent, and we can determine the term. We find that the exponent $\gamma \approx -1.5$. Currently, we do not understand the origin of such a non-integer power, and since it is determined numerically, it might just be one among further subleading terms. Furthermore, it is unclear if it is linked to the topology of space since, as mentioned, we cannot determine it reliably for spherical topology. {Of course, such terms are of great interest since they might be important when the spatial volume of the universe is small. We hope to be able to determine the terms more precisely in future simulations.} 

It is natural to compare the action \rf{j20} to the standard Hartle--Hawking minisuperspace action in spaces with Euclidean signature,
which is derived from the standard Einstein--Hilbert action by assuming isotropy and homogeneity of space.
If we fix the time coordinate to be proper-time with a lapse function equal 1, the minisuperspace action can be written as 
\begin{equation}\label{eq:scovcont}
S[v] = \int d t \, \left[ \frac{1}{\tilde\Gamma} \frac{\dot{v}^2}{v} + \tilde\a \, v^{1/3} + \tilde\lambda \, vs. \right], 
\end{equation}
where $v(t)$ is the physical spatial volume. Here, $\tilde{\Gamma}$ is negative and $\a\! =\! 0$  for $T^3$, while for the sphere, there is a precise relation between $\tilde{\Gamma}$ and $\a$ (see \cite{impact-topology} for a detailed derivation). The negative sign of the ``kinetic'' term reflects the unboundedness of the Euclidean Einstein--Hilbert action. We are using a discretized version of the Einstein--Hilbert action, and our action is also unbounded from below in the $N_4 \to \infty$ limit. However, a remarkable interplay between the entropy of configurations with the same value of the action and the actual value of the discretized action seems  to be able to cure the unboundedness problem and produce a kinetic term with a healthy sign. It is also remarkable that despite the negative sign of the ``kinetic'' term in \rf{eq:scovcont}, the extremum of $S[v]$, found under the constraint that the total four-volume is fixed, i.e., that $\tilde{\lambda}$ is a Lagrange multiplier, produces precisely the solution \rf{ja3} with $\la n_t \ra$ replaced by the solution $\bar{v}(t)$ to \rf{eq:scovcont}.

\section{Semiclassical Geometry and Baby Universes}\label{sec4}

\subsection{Boundaries and Elementary Cells}

As mentioned above, in some sense, the DT formalism contains GR in its purest form: we only know the local neighborhood, the vertex where we are and its connection to the neighbors, and we are equipped with our measuring devices, the links, which provide us with the geodesic distance to the neighbors, and from there we work our way out and determine the geometry. This is to a large extent what we have done so far in the computer simulations, both in the old DT studies and in the spatial directions in CDT. However, the description one obtains this way has its limitations, in particular if we want to understand some characteristic features of a typical geometry appearing in the path integral. While a configuration (a geometry) picked from the path integral is not physical, in the same sense that a path of a particle in the path integral is not physical (we cannot measure it), it might nevertheless tell us a lot about how we should think about the quantum aspects of geometry. Here, a good coordinate system can be useful. The time coordinate in CDT is an example where a ``good'' choice of a coordinate allows us to determine an effective minisuperspace description of our quantum universe. In the case where the topology of the spatial slices is that of $S^3$, we have not been able to find equally useful coordinates in the spatial directions. However if the topology of the spatial slices is that of $T^3$, one can use the periodic structure,  as well as the non-contractible loops of $T^3$, to obtain new information about the geometry of the path integral configurations. 

The first thing to be aware of is that the computer simulations change the connectivity of the triangulations. While the change by construction respects the time foliation, there is no such restriction in the spatial directions. If the starting configuration was constructed with  $T^3$ topology, the local changes  of the triangulation made by the computer simulations will preserve this topology, but not any initial foliation of $T^3$. If we want to make maximal use of the periodic structure of  a triangulation with $T^3$ topology,  we have to identify such structure, and since it is in no way unique (even for a regular triangulation), we also want a ``reasonable''  periodic structure, which is a somewhat non-trivial task if the triangulation is quite fractal. Let us briefly describe how we define the periodic structure (for details, we refer to \cite{pseudo-car}). The idea is to start with a regular triangulation where the periodic structure can be defined in a trivial way (see again \cite{pseudo-car} for details) and then keep track of this structure and modify it appropriately, such that when the triangulation is changed in the Monte Carlo simulations, we can still identify a ``reasonable'' periodic structure despite having  quite a fractal geometry. The periodic structure will be defined by {boundaries}, and it is convenient to view $T^4$ as an infinite system of {elementary cells} in $\mathbb{R}^4$, separated by these boundaries. In order to keep track of the boundaries, we introduce a labeling of the   dual lattice. Each simplex in the triangulation is represented as a vertex in the dual lattice, and each tetrahedron, being the boundary between two simplices, is represented as a link $(i, j)$ between the two vertices $i$ and $j$ on the dual lattice. The concept of the spatial directions (and time direction) is {well} defined for our nice starting triangulation, which is a triangulation of a hypercube whose opposite sides are identified, {and thus we can use it to} define boundaries that are ``orthogonal'' to what we denote $x,y,z$ and $t$ directions, and where the $t$ direction will coincide with our CDT time direction. We label the three spatial boundaries $x,y$ and $z$, as we will try to define coordinates where the boundaries correspond to $x=0$, etc. We now assign four values  $B_{i j}^\sg$ (a four-vector, one can say) to each link $(i, j)$, where $\sg =x,y,z$ or $t$. The value  0 corresponds to the case when both $i$ and $j$ belong to the same cell.  Values $B_{i j}^\sg =\pm 1$ imply that the link $(i, j)$ crosses the $\sg$-boundary between two cells in the positive or in the negative direction. If we have a closed loop of dual links, defined by a sequence of vertices $i_n$, $n=1,\ldots,k$, where $i_{k+1} = i_1$ (up to the cell number), we may form the sum of vectors $\sum_{n=1}^k B^\sg_{i_n i_{n+1}}$  along the loop. For a contractible  loop, the sum must be the zero vector, while for a non-contractible loop (closed by the periodic boundary conditions), the sum gives  topological information about the winding number of the loop. The boundaries are not physical, and the Einstein--Hilbert action \rf{SRegge} does not depend on the boundaries. The computer algorithm changes one triangulation into another one by performing  local changes. When we perform such changes, we check if the boundaries between elementary cells pass through the affected region of the triangulation. If not, we perform the change of the triangulation as if the boundaries were not there. If a boundary between two cells crosses the affected region, we deform the boundary, but not the triangulation itself, in such a way that it is still connected but outside the region. We then perform the change of geometry, and we try to modify the boundary in the new geometry such that its size, i.e., its three-volume, will be minimized. From time to time during  the computer generated changes of the triangulation, we check if a local modification of boundaries can minimize the three-volumes of the boundaries in all directions. In this way, we try to keep the boundaries ``reasonably" small during the changes of the triangulation generated by the computer.

\subsection{Pseudo-Cartesian Coordinates}\label{Sec:pseudo}

\begin{figure}[t]
\begin{center}
{\includegraphics[width=13cm]{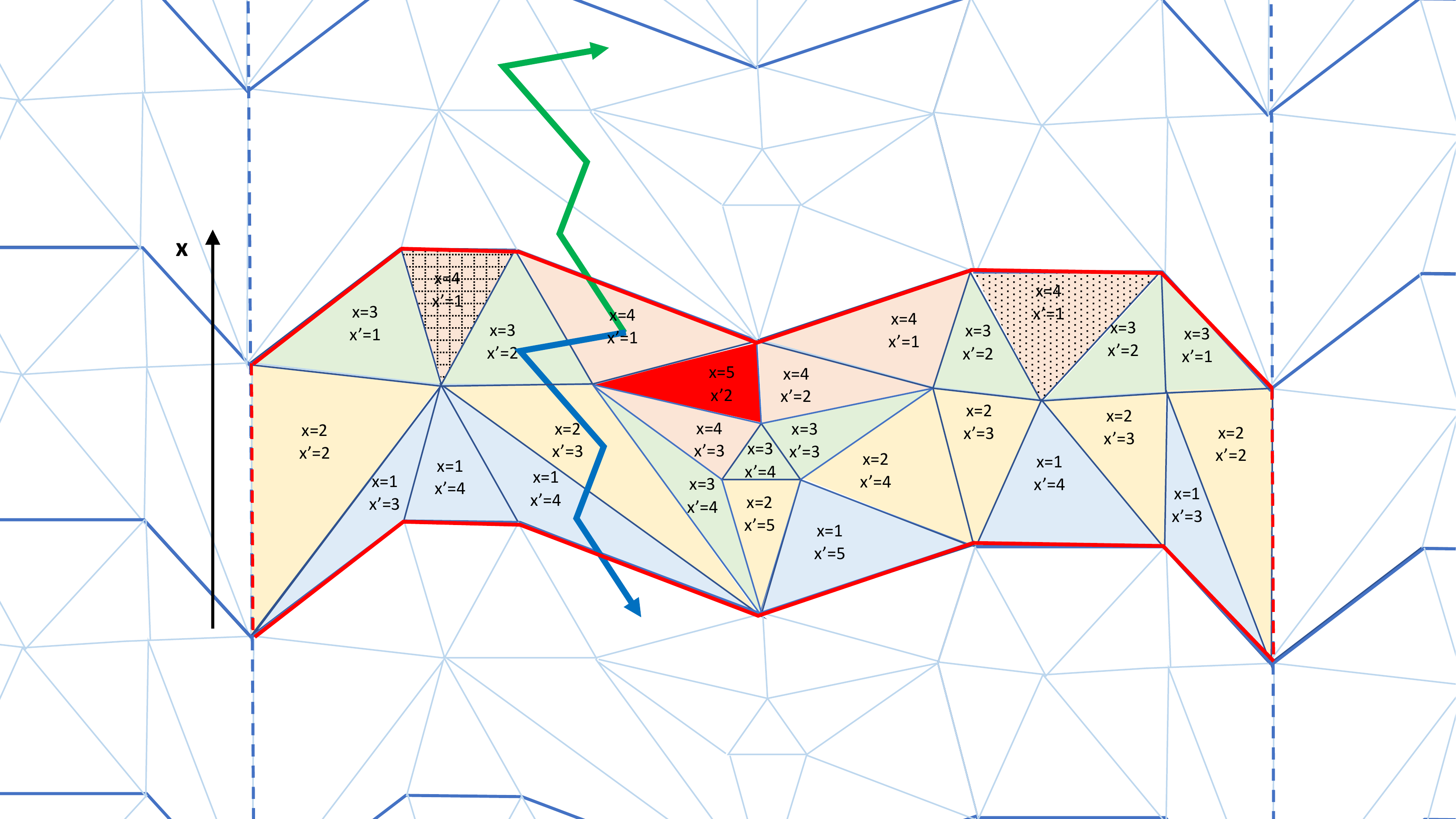}} 
\end{center}
\caption{{A (two-dimensional) visualization of a triangulation (colored triangles) with toroidal topology. The (smallest) boundary orthogonal to the   ${x}$ direction is plotted as a red solid line, and the boundary orthogonal to the other direction as a red dashed line (the red solid lines, as well as the red dashed lines, are identified). Different colors mark  different ${ x}$ coordinates. All triangles of the same color and texture  form single { slices}. For ${ x}= 1, 2, 3, 5$, one has only a single slice for each  ${x}$ coordinate, but for ${ x}=4$, one can distinguish three separate slices. All triangles but the dark red one belong to the trunk, and the dark red triangle belongs to a branch. For each triangle, one can find (one or more)  minimal loops with a nontrivial winding number: all triangles whose centers are marked by a solid zigzag arrow belong to the same $\{1,0\}$  loop (green arrow) and also to the same  $\{-1,0\}$ loop (blue arrow); the length of these loops is $4$.
} }
\label{fig:visualize}
\end{figure}

\begin{figure}[t]
\includegraphics[width=6.5cm]{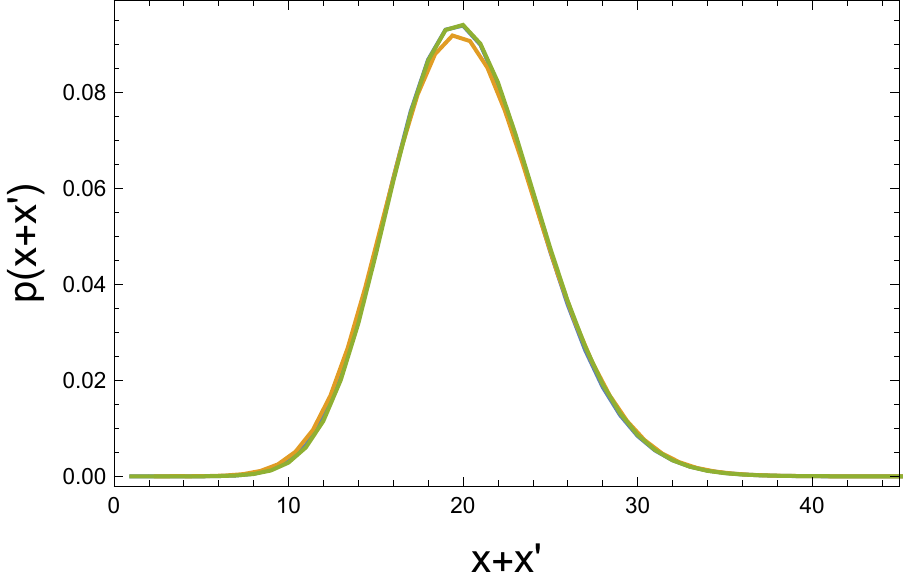} 
\includegraphics[width=6.5cm]{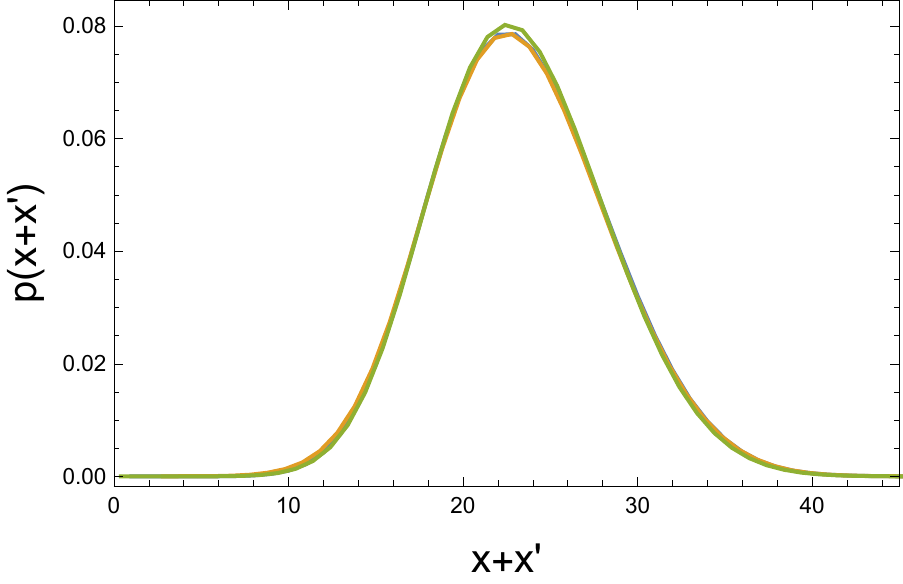} 
\caption{Distributions of $p({ x}+{ x'})$ (blue),  $p({ y}+{ y'})$ (green) and $p({ z}+{ z'})$ 
(orange) for systems with $N_{(4,1)}=80k$ (left) and $N_{(4,1)}=$160k (right) and $T =4$. One can show that 
the distributions scale as $ \big(N_{(4,1)}\big)^{1/4}$. }
\label{shifts}
\end{figure}

Once the boundaries of a toroidal triangulation are defined, they can be used to erect what we denote a {pseudo-Cartesian} coordinate system. Let us consider a spatial boundary, e.g., the $x$-boundary of an elementary cell.
For {every (dual lattice)} link $(i, j)$ that crosses the boundary with value $B^x_{ij}=+1$,
we assign to vertex $j$ a coordinate $x=1$.
From each already visited vertex, we move to the neighboring vertices, but we are not allowed to cross the boundary in the negative direction. We mark these vertices with coordinate $x=2$ and continue this way until each vertex in the dual graph has an $x$-coordinate. The way this is actually done is by studying a discretized version of a diffusion equation on the dual lattice; for details, we refer to \cite{pseudo-car}. It is also possible to cross the boundary in the negative direction, and we can assign a coordinate $x' =1$ to the vertices we meet on the negative side of the boundary and then extend this construction to each vertex, as for the $x$ coordinate. In this way, each vertex $v$ has two coordinates: $x(v)$ and $x'(v)$. The construction described is illustrated on  a two-dimensional lattice in Figure~\ref{fig:visualize}. 
Do these spatial coordinates qualify as good coordinates? We are mainly interested in this question for configurations in 
the de Sitter phase $C_{dS}$.
In Figure \ref{shifts}, we show the probability distribution $p(x(v)+x'(v) )$. 
For a completely regular triangulation, {$p(x+x')$ would be zero except for one peak observed for
$x+x'=h_x$}, the length of the  shortest non-contractible 
loop crossing the $x$-boundary. Clearly this is not what we observe, showing us that the elementary cell after many changes of the local geometry is no 
longer that regular. While this indeed is to be expected, unfortunately the coordinates are not that optimal. If we look at 
the set of points with  constant $x$, we would like this to be a connected hypersurface, but this is not the case.

For a detailed analysis, we refer to \cite{pseudo-car}, and here let us just describe the situation qualitatively: given a simplex, i.e., a vertex $v$ on the dual lattice,  with coordinate $x_0=x(v)$, we can identify  the vertices $\{w(v)\}$ with coordinates $x(w) =x_0$, where in addition $w$ is connected to $v$ by moving only between vertices $u$ where  $x(u) \geq x(v)$. We call such a set the {\it slice} associated with $v$. The set of all vertices with coordinate $x_0$ will in general consist of a set of disconnected slices. A slice at $x_0$ is connected via  links to precisely one slice at $x_0-1$ but can connect to many (or zero) slices at $x_0+1$. All vertices at $x_0=1$ are connected and thus  form a slice, which we denote the {\it root} of the {\it trunk}, where the trunk consists of the slices which are connected with slices of increasing $x$ all the way from the root to the ``top'' boundary of the cell. The trunk may split into several pieces when reaching the top boundary. Furthermore, starting at a slice at $x_0$ and following its connections to slices at $x_0+1$, etc., this set of slices might never reach the top boundary. We call such a set of slices a {\it branch}, and the branches clearly have a {tree structure}. All this is in principle illustrated in \mbox{Figure~\ref{fig:visualize}}, where the deep red triangle with coordinate labels $(x,x') = (5,2)$ is a branch. Figure~\ref{branches} shows, for a large triangulation, the number of branches one encounters at coordinate $x$ and their corresponding volumes. The number of branches is large, but their volumes are not that large. The existence of these branches and the corresponding disconnectedness of the hypersurfaces of constant $x$ implies that the pseudo-Cartesian coordinates are not an optimal choice of coordinates; in Section~\ref{sec5}, we will introduce a better set of coordinates. However, they highlight the fractal structure of a typical geometry, which seems to share a number of characteristics with the  fractal geometry encountered in 2d DT  \cite{kawai-wata,aw,ajw} and in the so-called generalized 2d CDT \cite{cdtsft,stochastic-cdt}, something definitely worth exploring. It turns out that we can dissect the fractal structure slightly further by using the non-contractible loops introduced and the periodic structure of the triangulations. We now turn to this~topic.

\subsection{ Semiclassical and Fractal Geometry from  Loops}

We now have the setup to use non-contractible loops to obtain information about the toroidal geometry of a generic CDT configuration. 
{Let us consider the loops passing through a vertex $v$ in the 
dual lattice, i.e., a given simplex in the triangulation.
The loop  lengths 
of  the shortest non-contractible loops around the torus in what we have called the $x$, $y$ and $z$ directions 
tell us much about the position of the vertex $v$ in the torus. 
The logic behind this is illustrated in Figure~\ref{fignew}. If the torus consists of a ``semiclassical'' center and ``quantum'' outgrowths, the loop lengths will be large when $v$ is in the outgrowths and smaller when $v$ is in the center. We will now present some evidence for this picture. 
\begin{figure}[t]
\begin{center}
\includegraphics[width=7cm]{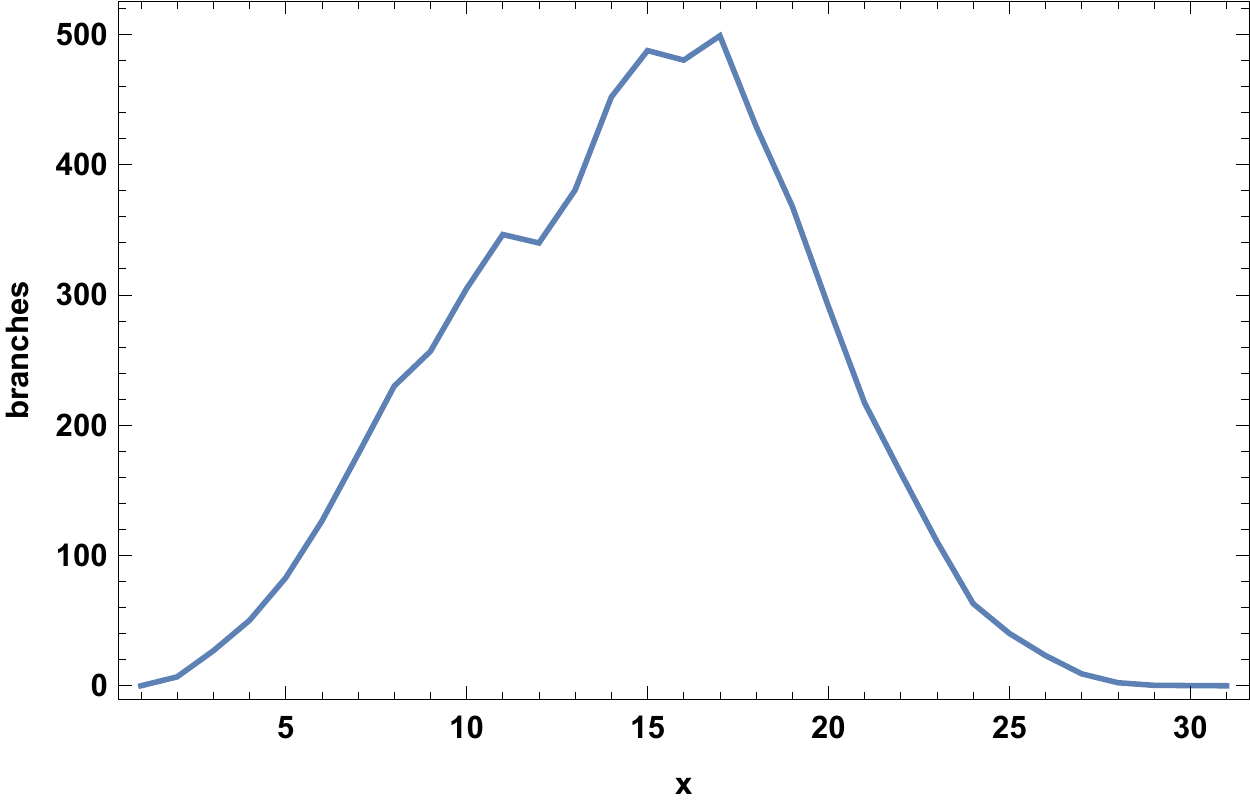}~~
\includegraphics[width=7cm]{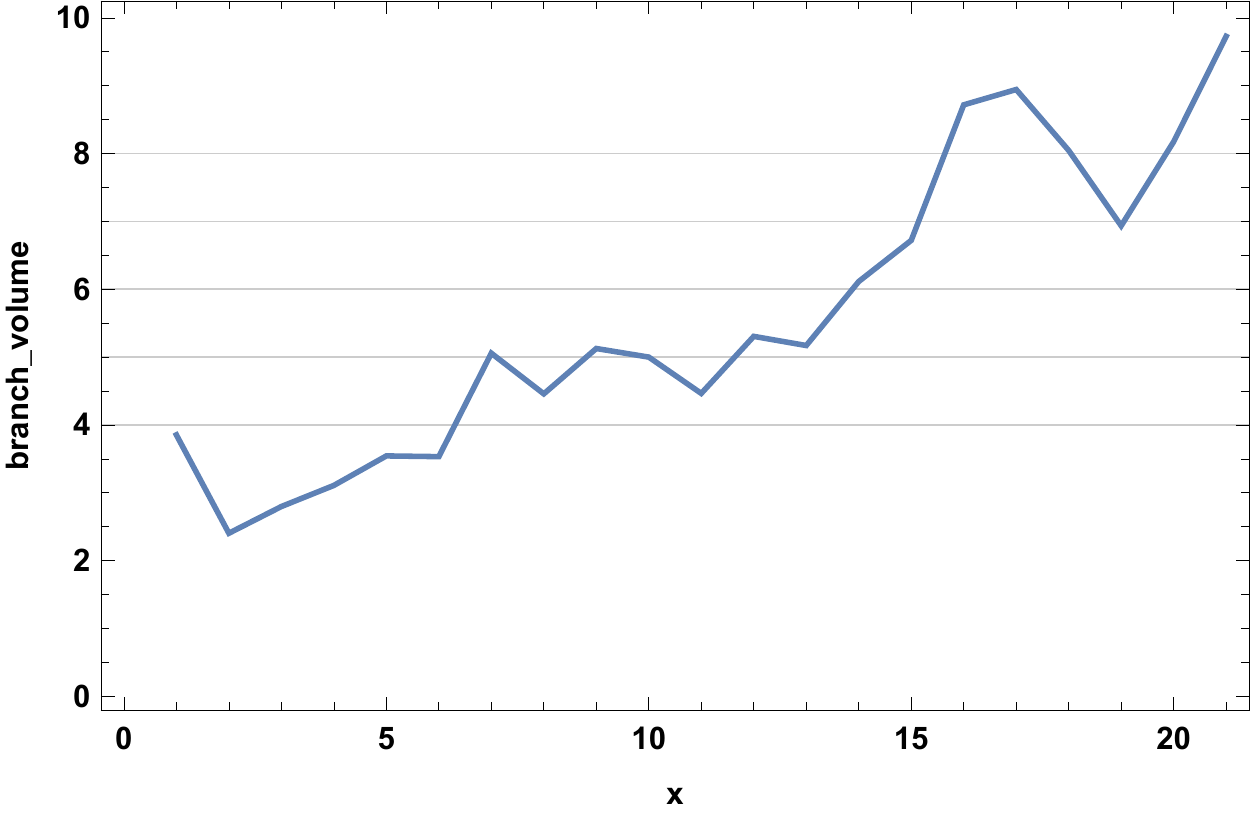}
\end{center}
\vspace{-0.5cm}
\caption{ The number of branches at a distance ${x}$ (left figure) and the average branch volume (right figure) in a triangulation from phase $C_{dS}$. } 
\label{branches}  
\end{figure}

\begin{figure}[t]
\includegraphics[height=55mm]{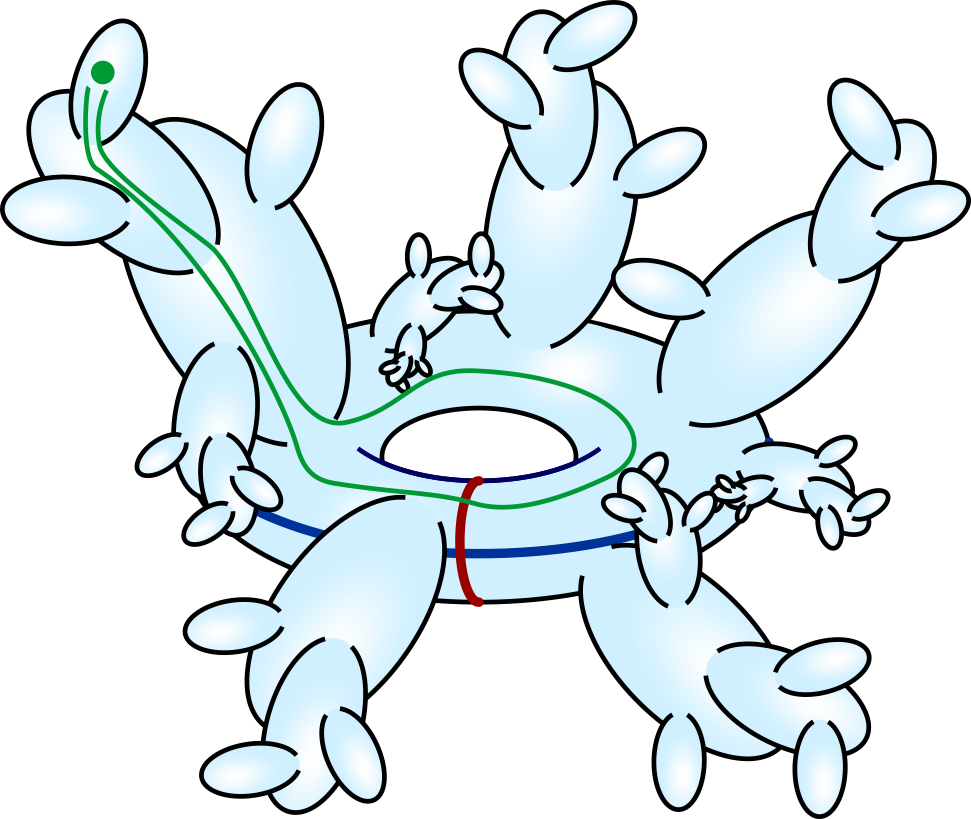}
\hspace{0.8mm}
\includegraphics[height=55mm]{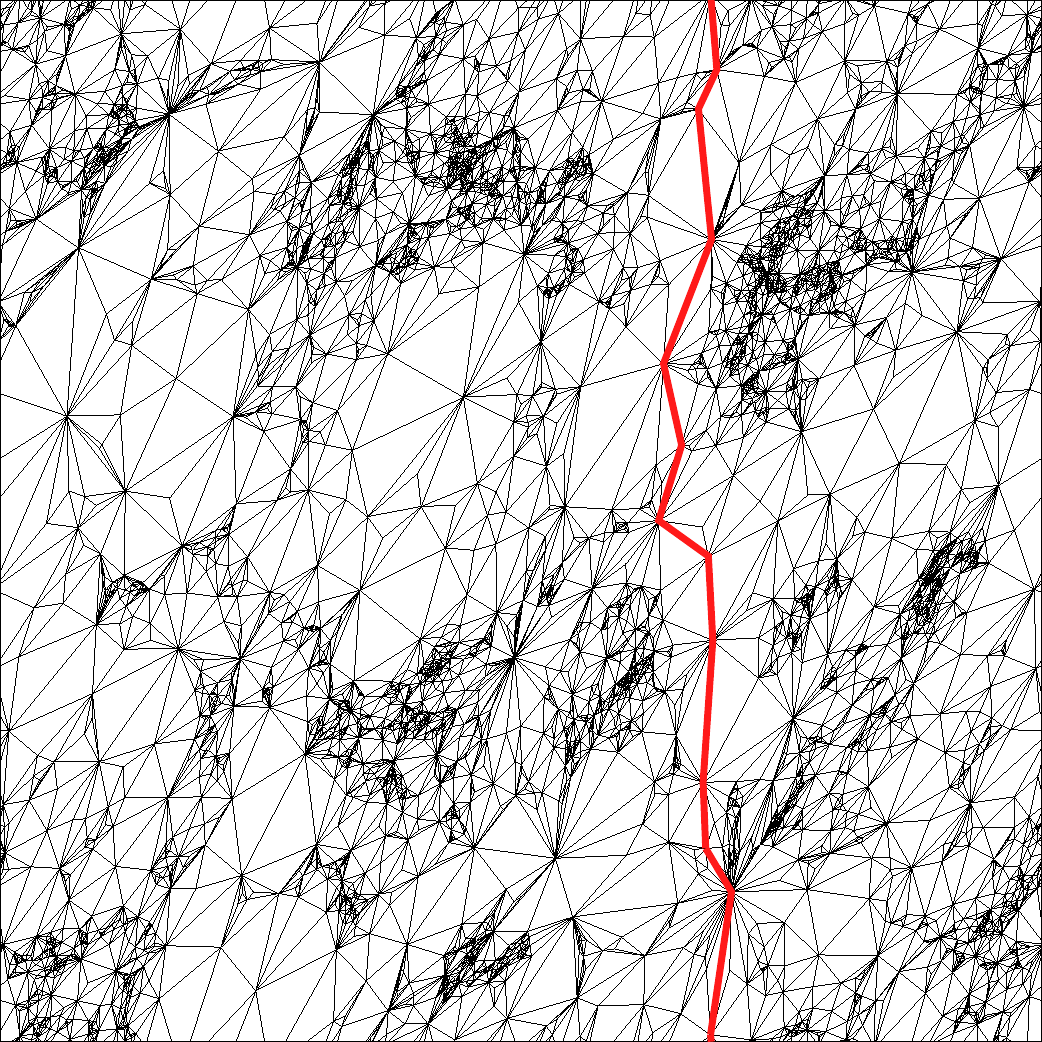}
\caption{
Left: illustration of a torus with outgrowths. 
The blue and red lines represent two non-equivalent and non-contractible loops.
The green loop is the shortest loop passing through the green point in the same direction as the blue line. Right: embedding of a triangulation of the two-torus consisting of 150,000 triangles into the Euclidean plane (picture from \cite{ab}).
Shown in red is the shortest non-contractible loop.}
\label{fignew}
\end{figure}

{According to our discussion}, the shortest loops should wind around the center of the torus. Thus, it is convenient to denote the length of the shortest non-contractible loops of a vertex $v$ as {height} labels $h_x(v),h_y(v),h_z(v),h_t(v)$, telling us how far inside an outgrowth $v$ is. Evidence for such an interpretation can be found by using the representation of the torus $T^4$ as a repetition of elementary cells on $\mathbb{R}^4$. On $\mathbb{R}^4$, each copy of the elementary cell is labeled by ``cell coordinates'' $(n_x,n_y,n_z,n_t)$, constructed from the  number of times one passes the cell boundaries from the starting elementary cell. By studying the above-mentioned diffusion equation on $\mathbb{R}^4$, we can then find the shortest path between copies of a vertex $v$ in the other cells.
{The length of  such a 
path is not necessarily just a multiple of the length of the shortest path between $v$ in neighboring cells,}
the reason being that we can increase the winding number of a {loop going through a} vertex in an outgrowth by first moving to the central part and then circling a number of times around the center before finally moving out in the outgrowth again in order to close the loop. This is illustrated in Figure~\ref{fig:av}. For the particular configuration used in  the figure, the shortest loop in the x-direction is 18.  Furthermore, it turns out that for this configuration, there is only one loop of length,  18. Rather remarkably, if we we ask for the shortest loop winding eight times around the torus in the $x$-direction, it will wind around at least one time following precisely this shortest loop, no matter 
from which vertex $v$ we start.  

\begin{figure}[t]
\centering
\includegraphics[width=10cm]{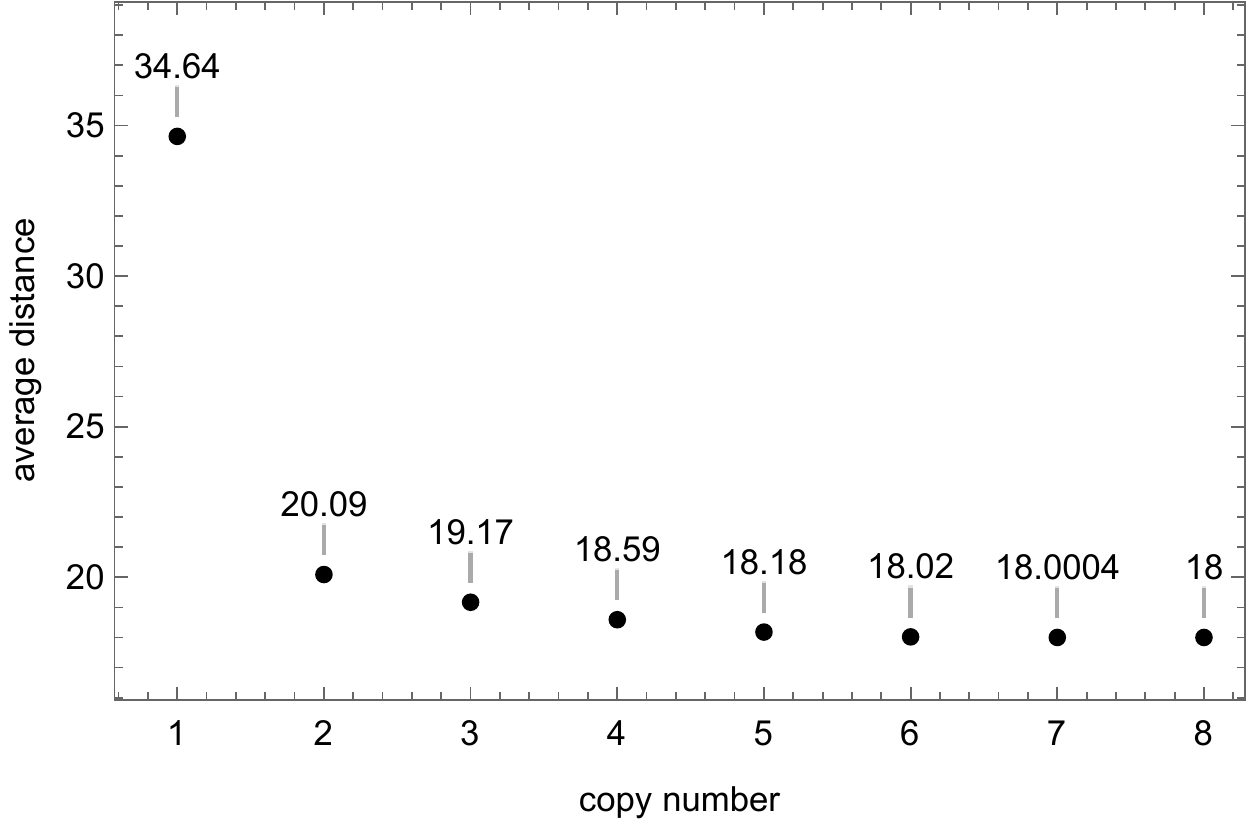}
\caption{Distance from a starting simplex to its copy in cell $\{n,0,0,0\}$ minus distance from the same simplex to its copy in cell $\{n-1,0,0,0\}$, averaged over all the simplices of the configuration. Already for $n=8$, the minimal value of 18 is reached.}
\label{fig:av}
\end{figure}

As shown in Figure~\ref{fig:sim2}, we have chosen a vertex $v_0$ deep into an outgrowth with a large height $h_x(v_0)$. We then follow the vertices $v$ that belong to  the shortest loops associated with $v_0$ when the loops wind around the torus one, two, three and six times. It is seen that $h_x(v)$ decreases when $v$ approaches the center. At some point, $v$ starts to belong to the shortest loop of length 18, and it stays there, i.e., $h_x(v) = 18$,  winding around a number of times, until it leaves and moves into the outgrowth again in order to return to $v_0$ and close the loop. When it returns to the outgrowth, $h_x(v)$ starts to increase again, until eventually it reaches the value $h_x(v_0)$. It is also seen that the values of $h_y(v)$ (and $h_z(v)$) are quite correlated with the values $h_x(v)$. Heights in the outgrowths are in general large in all directions, and all heights in the center 
are small.  
   
\begin{figure}[t]
\centering
\begin{tabular}{cc}
\includegraphics[width= 0.47 \textwidth]{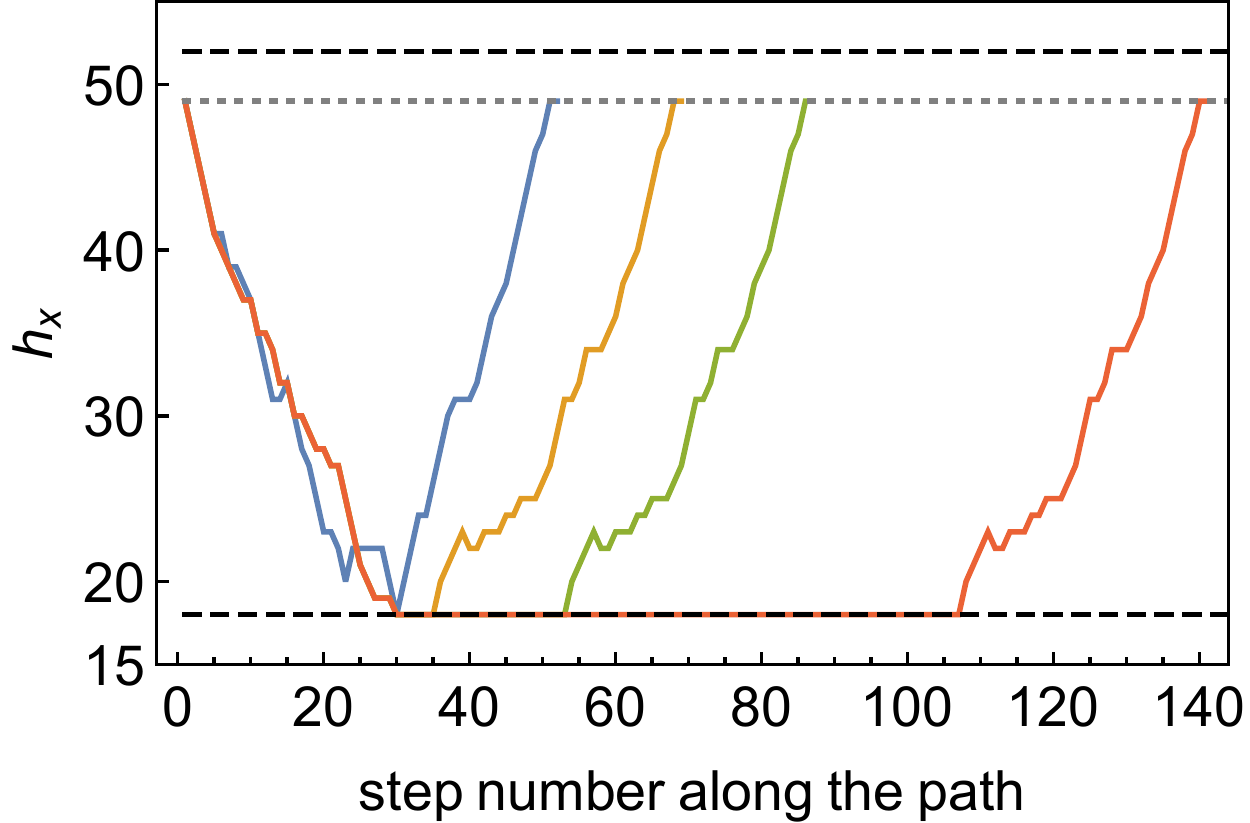} &
    \includegraphics[width= 0.47 \textwidth]{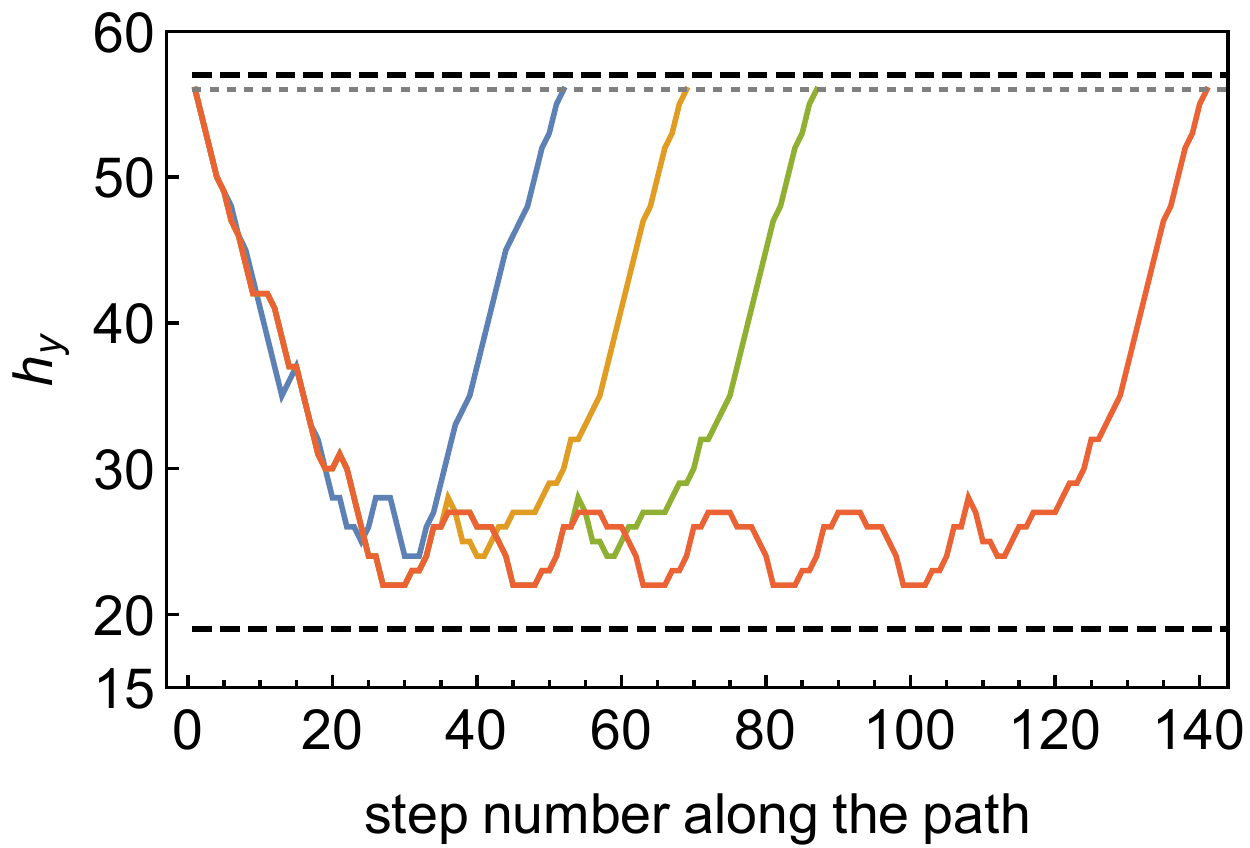}
\end{tabular}
\caption{Heights in 2 basic directions of consecutive simplices along loops of winding
numbers $\{1,0,0,0\}$ (blue), $\{2,0,0,0\}$ (orange), $\{3,0,0,0\}$ (green), $\{6,0,0,0\}$ (red) starting from a simplex in an outgrowth. The dashed lines indicate the minimal and maximal heights in the configuration, and the dotted line indicates the height of the initial simplex.}
\label{fig:sim2}
\end{figure}

A number of other measurements, for which we refer the reader to \cite{fractal-torus}, corroborate the picture advocated above, namely that in the de Sitter phase of the toroidal CDT the spatial $T^3$ part consists of a relatively small toroidal center, with outgrowths of almost spherical topology which contain most of the simplices. Measurements show that the number of simplices (vertices in the dual graph) where the height function has a local maximum is not small. This suggests that the outgrowths, even if they have almost spherical topology (except for the small region of attachment to the center), are quite fractal, having a large number of baby universes. {Furthermore, in the DT models of gravity baby universes have played an important role in determining the fractal structure of the geometries} \cite{babyuniverses,Ambjorn:1993nr,Ambjorn:1993vz}}. An important feature of the non-contractible loops is that the length distribution scales as $N_4^{1/4}$, as {also noted for the distribution of $x+x'$} in the figure caption to Figure~\ref{shifts}. The consequence of such a ``canonical'' scaling is that the volume of the toroidal center, although small compared to the volume of the outgrowths,  will also scale canonically.  This justifies thinking of it as semi-classical. This is  in contrast to the toroidal center part of the two-dimensional configuration shown in the right  part of  Figure~\ref{fignew}, which can be shown to vanish in the large volume limit, and thus is not semiclassical but entirely quantum, a known feature
of 2D DT.

\section{Toroidal Coordinates Via Scalar Fields}\label{sec5}

In the pseudo-Cartesian coordinates approach, {discussed in Section \ref{Sec:pseudo} above,} the 
constant-coordinate hypersurfaces were quite disconnected. This can be traced back to the fractal structure of the configurations, when viewed in terms of geodesic distances. The pseudo-Cartesian coordinates were basically labeled by their geodesic distance to the cell boundaries, and if there is a fractal structure when expressed in terms of geodesic distances, it is unsurprising that the distance function from a cell boundary will have many local maxima, i.e., the {branches} that we encountered. How can we do better? What comes to mind is that harmonic functions, solved on compact domains with suitable boundary conditions, never have local extrema in the interior of the domain. It thus seems that if suitable boundary conditions are used for our elementary cell, we might use a solution to the Laplace equation as a coordinate, increasing when moving from the lower to the upper boundary of an elementary cell. Let us now show that this works, and let us first discuss the problem using the standard language of differential geometry and then show that it can be translated to our triangulations. 

Consider two Riemannian manifolds $\mathcal{M}(g_{\mu\nu})$ and $\mathcal{N}(h_{\alpha\beta})$, with metrics $g_{\mu\nu}(\xi)$ and $h_{\alpha\beta}(\eta)$, where $\xi$ and $\eta$ denote some coordinates on  $\mathcal{M}$ and $\mathcal{N}$. A harmonic map $\phi^\sg$ from  $\mathcal{M}$  to $\mathcal{N}$ can be defined as a map that minimizes the following action:
 \begin{eqnarray}
S_M[\phi]	= \frac{1}{2} \int \mathrm{d}^n \xi \sqrt{g(\xi)} \; g^{\mu \nu} (\xi) \; h_{\rho \sigma}(\phi(\xi)) 
\; \partial_\mu \phi^\rho(\xi) \partial_\nu \phi^\sigma(\xi).
\label{classical_field_eq}
 \end{eqnarray}
  In our case, we are interested in the situation where   $\mathcal{M}$ has the topology of $T^4$  and $\mathcal{N}$ the topology of $T^1 \equiv S^1$.  Since  $\mathcal{N}$ is one-dimensional, we can always choose $h_{\rho \sigma}=1$, and the Laplace 
 equation reads
 \begin{eqnarray}
\Delta_\xi \phi(\xi) = 0,\quad 
\Delta_\xi = \frac{1}{\sqrt{g(\xi)}}\frac{\partial}{\partial \xi_\mu} \sqrt{g(\xi)}\,g^{\mu\nu}(\xi)\frac{\partial}{\partial \xi_\nu}.
\label{laplace_eq}
 \end{eqnarray}
 
 Let us for illustration first consider the case where   $\mathcal{M}$ also has the topology of $S^1$. In that case, we can also choose coordinates $\xi$ such that $g_{\m\n}(\xi)$ is constant, but we prefer to keep the general notation.  We want  $\xi \to \phi(\xi)$ to be a non-trivial map $S^1\to S^1$ such that $\phi$ can serve as a coordinate instead of $\xi$. One way to implement this is to represent $S^1$ periodically on $\mathbb{R}$, precisely as we repeated the elementary cell periodically on $\mathbb{R}^4$, and require that $\phi$ satisfies
\begin{equation}
\phi(\xi+n)=\phi(\xi)+n \delta,    
\label{phi_cont_eq}
\end{equation}
which maps the circle with a unit circumference to a circle with a circumference $\delta$. The solution to the Laplace equation in this case satisfies
\begin{equation}\label{jx1}
 \mathrm{d} \phi(\xi) = \delta \cdot \sqrt{g(\xi)} \; \mathrm{d}\xi.
\end{equation}
The solution $\phi(\xi)$ is fixed by choosing  a $\xi_0$ where $\phi(\xi_0)=0$. The map $\xi \to \phi(\xi)$ becomes a monotonically increasing invertible map in the whole $\mathbb{R}$, and we can use $\phi$ as a coordinate instead of $\xi$. With such a change of coordinates, the volume is invariant, i.e., $\sqrt{ g(\phi)} \,\mathrm{d}\phi = \delta \sqrt{g(\xi)}\,\mathrm{d}\xi$, or from \rf{jx1}, $g(\phi) = 1$. We can also consider the function $\psi(\xi) =  \textrm{mod}(\phi(\xi)-\phi(\xi_1),\delta)$. This satisfies the Laplace equation in the range between $\xi_1$ (where $\psi(\xi)=0$) and $\xi_1+1$ (where $\psi(\xi)=\delta$). The equation satisfied by $\psi(\xi)$ becomes a Poisson equation with the extra inhomogeneous local term, producing jumps at boundary points $\xi = \xi_1$ and $\xi=\xi_1+1$. It can still be considered to be a Laplace equation with a non-trivial boundary ``jump'' condition.  Generalizing this to  ${\cal M}$ with the topology of $T^4$, we want a solution to the Laplace Equation~(\ref{laplace_eq}) that wraps around $S^1$ in a particular direction once, and, in addition, we want the points $\xi$ in ${\cal M}$ that satisfy $\phi(\xi) =c$ to form hypersurfaces $H(c)$ whose union for $c$ varying in a range of length 1 covers the whole ${\cal M}$. Finally we want four such functions $\phi^\sg (\xi)$ and $\psi^\sg(\xi)$, where $\sg$ signifies the four chosen coordinate directions on ${\cal M}$, which we will denote $x,y,z,t$.

We now translate this formalism to our CDT triangulations. Let $N_4$ be the number of simplices in the triangulation, i.e., the number of vertices in the dual lattice. We label the vertices (or simplices) $i$.
In order to identify the boundaries, in Section~\ref{sec4}, we have defined the quantities $B^{\sg}_{ij}$ for links $(i, j)$  on the dual lattice.
We now extend the definition to include all pairs of vertices, in this way making $B^\sg_{ij}$ a (sparse)  $N_4\times N_4$ anti-symmetric matrix for each direction $\sg$ ($\sg =x,y,z$ and $t$). Thus, we have 
\begin{equation}
\label{adjacency}
{B}_{ij}^\sigma =
\begin{cases}
	\pm 1 & \textrm{if link $(i, j)$ crosses the boundary,}\\
	0 & \textrm{otherwise}.
\end{cases}
\end{equation}
The number of directed boundary faces of a simplex $i$ is given by $b_i^\sigma=\sum_j {B}_{ij}^\sigma$ and $\sum_i b_i^\sigma = 0$. For any simplex $i$ adjacent to a boundary, the values ${B}_{ij}^\sigma$ are all positive or zero (on one side of the boundary), or all negative or zero (on the other side). 
We consider four scalar fields $\phi_i^\sigma$ located in the centers of simplices, i.e., at vertices of the dual lattice, and solve the minimization problem for the following discrete version of the continuous action in Equation~(\ref{classical_field_eq}), for each field 
$\phi_i^\sigma$:
\begin{eqnarray}
\label{CDT_laplace_eq}
S_M^{CDT}[\phi^\sigma,{T}]= \frac{1}{2} 
\sum_{ (i, j)} (\phi_i^\sigma - \phi_j^\sigma -\delta \,{B}_{ij}^\sigma)^2 .
\end{eqnarray}
In (\ref{CDT_laplace_eq}),  the sum is over all links $(i, j)$ on the graph dual to  the triangulation $T$ (representing  
the manifold $\mathcal{M}(g_{\mu\nu})$ in Equation~(\ref{classical_field_eq})).
The parameter $\delta$ plays the same role as in the one-dimensional example considered previously. By rescaling the field, we can always set $\delta=1$, and we do that in the following. The action (\ref{CDT_laplace_eq}) is invariant under a constant shift of the scalar field  (the Laplacian zero mode). Furthermore, the boundary is immaterial if we consider a harmonic map $\phi$ from $T^4$ to $S^1$, and it was only introduced to represent $T^4$ and $S^1$ as $\mathbb{R}^4$ and $\mathbb{R}$. The manifestation of this is that the action stays invariant under a modification of the boundary ${B}_{ij}^\sigma$ and a shift by $\pm 1$ (depending on the side of the boundary) of the field value in a simplex $i$ adjacent to the boundary, i.e., ``moving'' the simplex to the other side of the boundary and compensating for the change of the field in its center. After such a move, the number of faces belonging to the boundary will in general change.

\begin{figure}[t]
\includegraphics[width=0.45\textwidth]{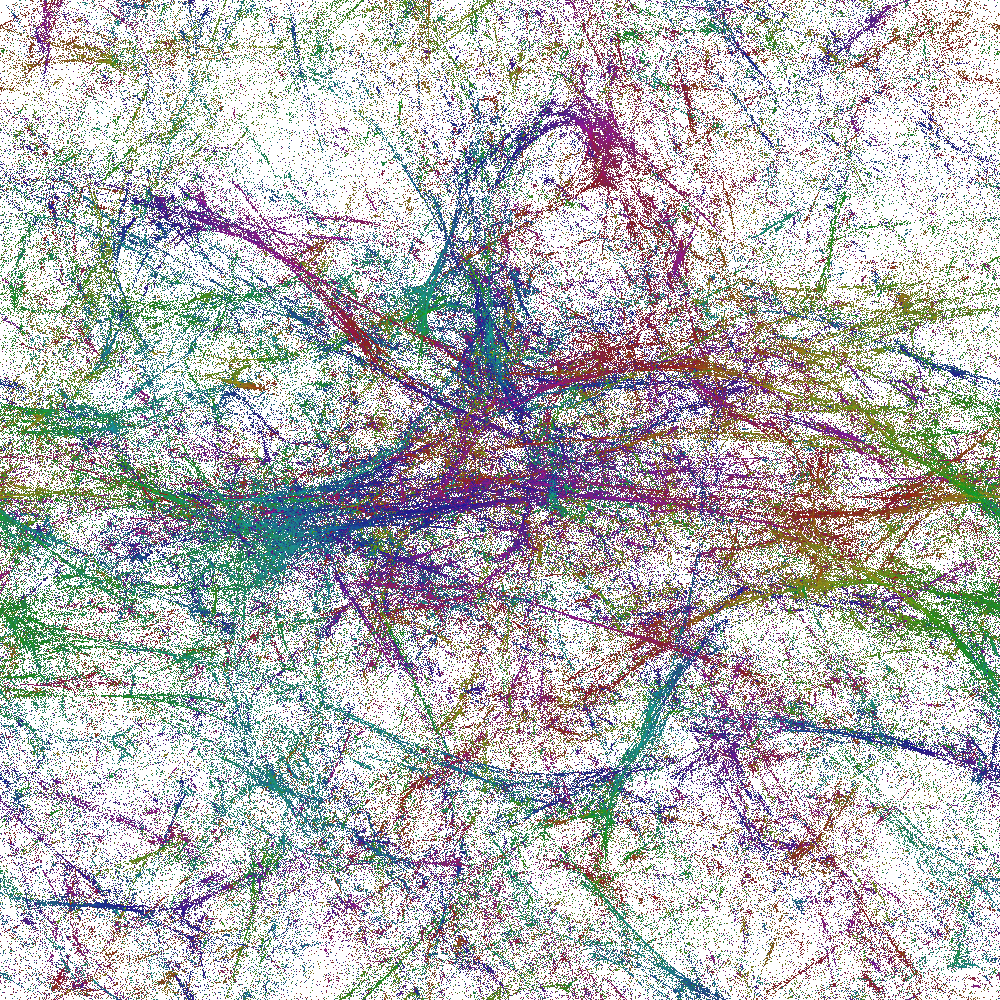} \hspace{0.5cm}
\includegraphics[width=0.45\textwidth]{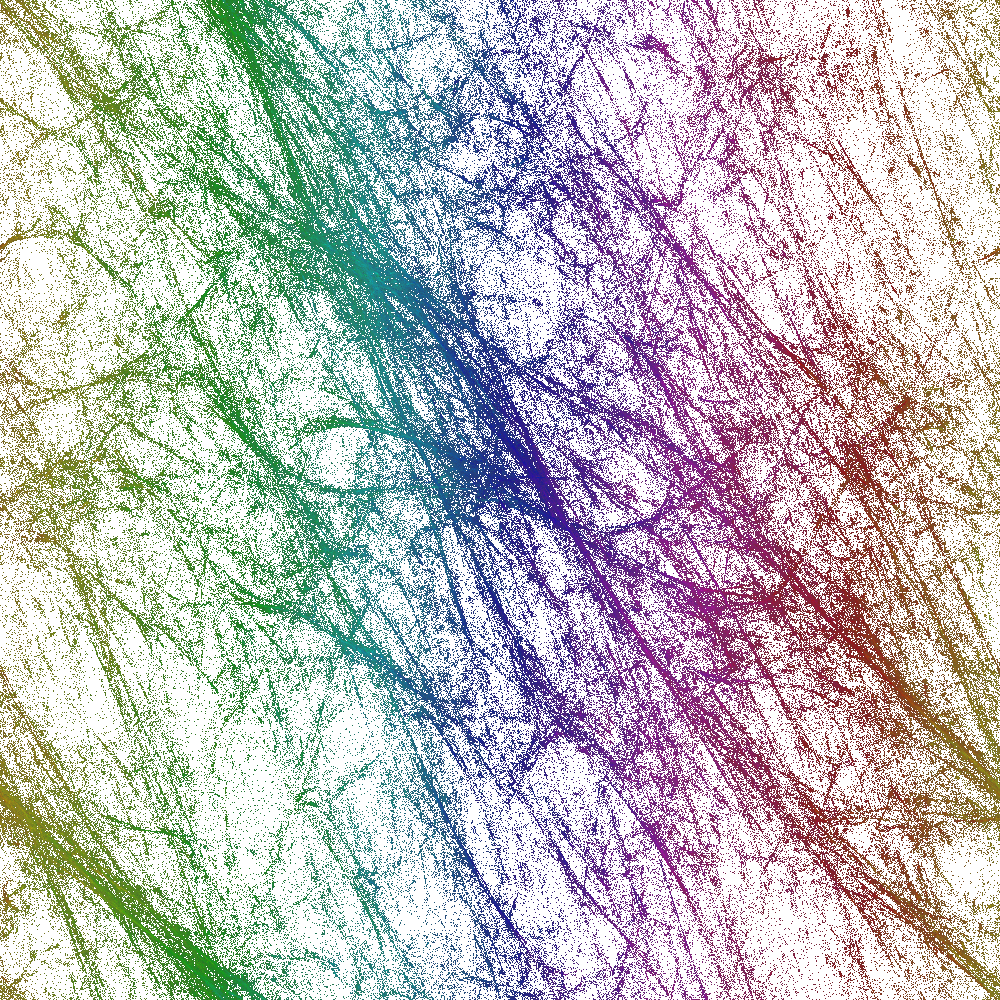}
\caption{Left figure is the  projection of four-volume on the $xy$-plane, 
as defined by (\ref{hypers_vol_eq_psi}) for a CDT configuration.
 Right figure is the projection on the $tx$-plane  of the same configuration. Different colors correspond to different times $t$ of the original $t$-foliation. 
There is a strong correlation between the original $t$-foliation (color) and new time coordinate $\psi^t$ (horizontal axis),
for the projection on the $tx$-plane.}
\label{phi-xy_plot}
\end{figure}

Let us now denote the classical field that minimizes the action (\ref{CDT_laplace_eq})  $\phi_i^\sigma$.
It satisfies the non-homogeneous Poisson-like equation
\begin{equation}
\label{classical}
    {L}\phi^\sigma =  b^\sigma,
\end{equation} 
where ${L}=5 { I}- {A}$ is the $N_4\times N_4$ Laplacian matrix and ${A}_{ij}$ is the adjacency matrix with entries of value 1 if simplices $i$ and $j$ are neighbors and 0 otherwise.
To be more precise, the Laplacian zero mode can be eliminated if we fix a value of the field $\phi_{i_0}^\sigma=0$ for an arbitrary simplex $i_0$. After that, we can solve for $\phi^\sg$ in \rf{classical}.
The analogue of the one-dimensional function $\psi(x)$ is given by 
$\psi_i^\sigma$:
\begin{equation}
    \psi_i^\sigma  = \textrm{mod}(\phi_i^\sigma, 1).
\end{equation}
As in the one-dimensional case, this definition eliminates the original boundaries but creates new ones defined by viewing 
\rf{classical} as an equation that defines the boundaries if we are  given a solution, i.e., the new boundary is defined 
by $\bar{b}^\sigma = {L}\psi^\sigma$. This allows us to reconstruct 
a new three-dimensional hypersurface $H$, separating the elementary cell from its copies in the direction $\sigma$ and characterized by the fact that the field jumps from 0 to 1 when crossing $H$. This hypersurface can be moved to another position if we consider a family of hypersurfaces $H(\alpha^\sigma)$ obtained from
\begin{equation}
\label{alpha}
    \psi_i^\sigma(\alpha^\sigma) = \textrm{mod}(\phi_i^\sigma-\alpha^\sigma, 1),\quad
      \bar{b}^\sigma(\alpha^\sigma) = {L}\psi^\sigma(\alpha^\sigma).
\end{equation}
By changing $0\leq \alpha^\sigma<1$, we shift the position of the hypersurface $H(\a^\sg =0)$ to that of $H(\a^\sg)$ and cover in this way the whole elementary cell. There are a number of subtleties associated with this since we have a discrete lattice structure and $\a^\sg$ is a continuous variable. This  implies that the various hypersurfaces defined in this way might  not be disjunct if two values  of $\a^\sg$ are too close. We refer to \cite{long-article} for details, and for the discussion of other non-trivial aspects of these hypersurfaces, but viewed in the right way, we can say that in this way we obtain a foliation in the direction $\sigma$. We may now use $\psi_i^\sigma= \psi_i^\sigma(0)$ as a coordinate in the $\sigma$-direction, and the same construction can be repeated in all directions $\sigma$. {Note also that the solutions viewed as belonging to $S^1$ are independent of precise locations of the hypersurfaces, as long as the deformations of the hypersurfaces respect the toroidal topology of the piecewise linear manifold}.  {Thus, we have obtained our goal:} for any toroidal configuration obtained in the numerical simulations, we can assign a set of coordinates $(\psi^x_i,\psi^y_i,\psi^z_i,\psi^t_i)$, all in the range between 0 and 1. In addition, these coordinates ``preserve'' the structure of the triangulation, since a  solution to the Laplace equation has the property that coordinates of each simplex are equal to the  mean value of coordinates of its neighbors. {The coordinates are the lattice analogy of harmonic coordinates, and harmonic coordinates are viewed as being as close as one can get in GR  to Cartesian coordinates representing an inertial system in special relativity. They are thus expected to be good coordinates in which to represent physical observables.} The coordinate $\psi^t_i$ is not the same as the one coming from the original foliation of the CDT model. However, they are quite correlated, as shown in Figure~\ref{phi-xy_plot}. Using our new coordinates, we can analyze the distribution of the four-volume (the number of simplices) contained in hypercubic blocks with sizes $(\Delta\psi^x_i, \Delta\psi^y_i, \Delta\psi^z_i, \Delta\psi^t_i)$, which is equivalent to measuring the integrated $\sqrt{g(\psi)}$:
\begin{eqnarray}
\Delta N(\psi) = \sqrt{g(\psi)} \prod_\sigma \Delta \psi^\sigma = N(\psi) \prod_\sigma \Delta \psi^\sigma.
\label{hypers_vol_eq_psi}
\end{eqnarray}

We can measure the full four-dimensional distribution $N(\psi)$, but to present it graphically, we show in Figure~\ref{phi-xy_plot} the projections of the volume density distribution of a typical configuration   on the $xy$-plane and the $tx$-plane, respectively, integrating over the remaining two directions. One observes a remarkable pattern of voids and filaments \cite{cosmic-void}, which qualitatively looks quite similar to pictures of voids and filaments observed in our real~Universe.


Using the new coordinates, we can now in principle repeat many of the measurements we originally performed using the CDT time coordinate. This includes studies of the profiles of the volume of hypersurfaces as a function of one of the $x,y$ or $z$ coordinates and  studies of the corresponding  volume--volume correlation functions and their minisuperspace actions.
However, we also have other options, since we now have reasonable coordinates in all directions,
as can be seen in Figure~\ref{phi-xy_plot}, which illustrates complicated structures at all scales.
All this is work in progress.

\section{Dynamical Scalar Fields with Non-Trivial  Topology}\label{sec6}

\begin{figure}[t]
\includegraphics[scale=0.23]{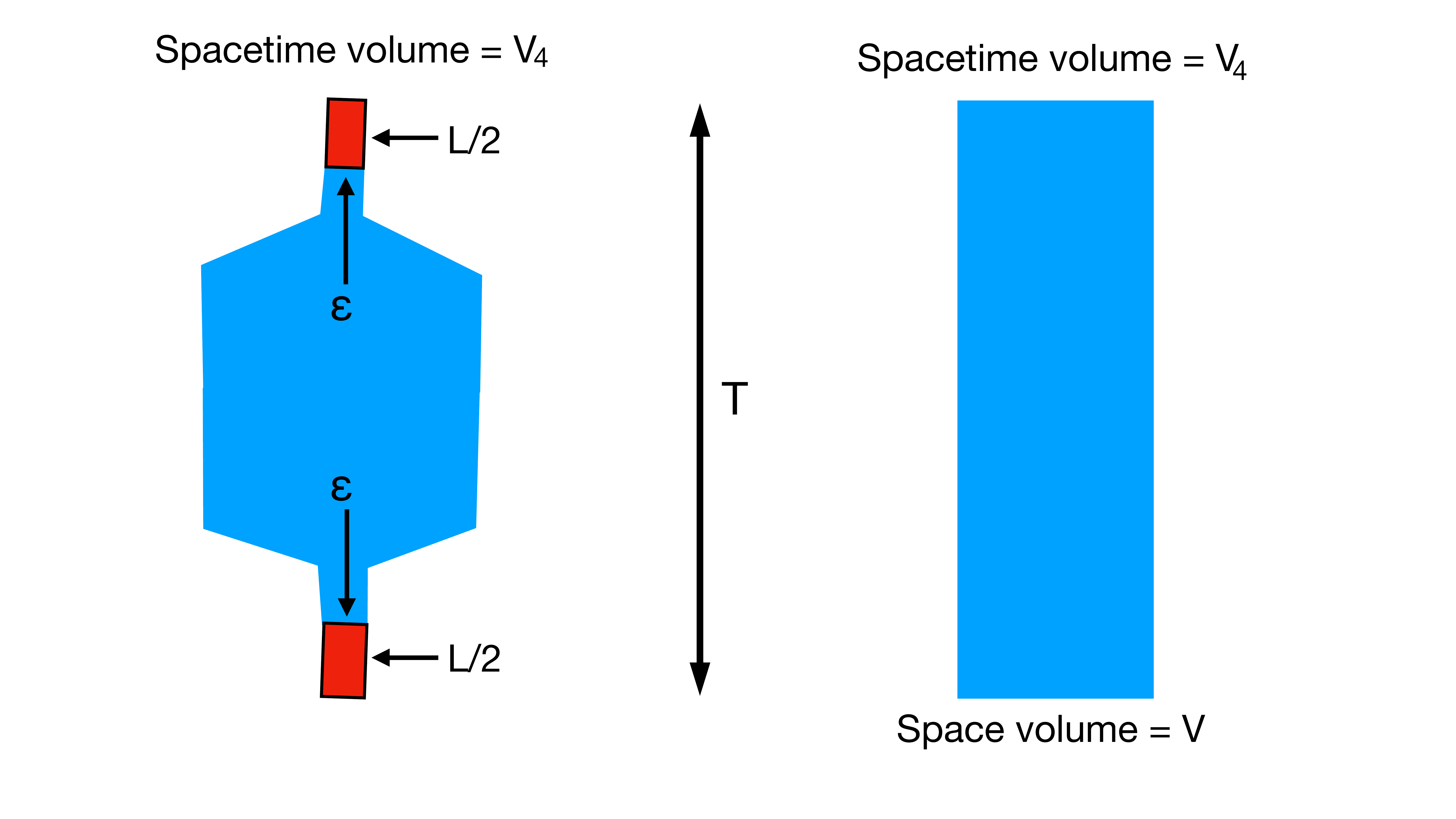}
\vspace{-1cm}
\caption{{Left}: a torus (opposite sides identified) with a pinch. 
{Right}:  a torus where $\phi$ is constant in horizontal direction and uniformly increases from 0 to $\delta$ from bottom to top. The two tori are assumed to have the same vertical length $T$ and the same spacetime volume $V_4$.}
\label{fig-pinch}
\end{figure}

Above we used classical  scalar fields with non-trivial winding numbers as coordinates for a generic quantum geometry configuration. Given the geometry, the fields were solutions to the Laplace equation.  We now ask what happens if such scalar fields, with action \rf{classical_field_eq}, are added to the path integral and become dynamical fields. As we will see, the interplay between the non-trivial topology of spacetime and the non-trivial topology of the scalar fields (the requirement that the fields have non-trivial winding numbers matching the non-contractible spacetime loops) will have a surprisingly dramatic effect on the geometry. Let us for simplicity consider the scalar field in the time direction, where the boundary conditions are simple when using the original CDT time coordinate, and where we, in the absence of the scalar field,  know that the volume profiles $v(t)$ are well described by the minisuperspace action \rf{eq:scovcont} with $\tilde\alpha =0$ when spacetime is toroidal. What we observed was that, to a good approximation, we had a solution $v(t) = V_4/T$ when the four-volume $V_4$ and the time extension $T$ were kept fixed. Adding a scalar field to the minisuperspace action, we obtain 
\beq\label{jax1}
\tilde{S}[v] = \int d t \, \left[ \frac{1}{\tilde\Gamma} \frac{\dot{v}^2}{v}  + \tilde\lambda \, vs. + v\, \dphi^2 +\kappa \dphi \right].
\end{equation}
Here we have assumed that the scalar field $\phi$ only depends on time, and in order to impose the constraint that $\phi[-T/2] = \phi[T/2] + \delta$, we have added the last term in \rf{jax1} with a new Lagrange multiplier $\kappa$, in addition to the Lagrange multiplier $\tilde{\lambda}$, which is there to ensure that the spacetime volume is $V_4$. One can now ask for the minimum of 
\beq\label{jax2}
S[v] = \int d t \, \left[ \frac{1}{\tilde\Gamma} \frac{\dot{v}^2}{v}  + v\, \dphi^2 \right]
\eeq
under the given constraints and find the solutions. We refer to \cite{long-article} for the details. However, the main points are easily explained without any analytic solutions. For small $\delta$, the constant solution is still the minimum with $\phi(t) = \delta\cdot  t/T$. However, for $\delta$ larger than some critical value $\delta_c$ of order 1,
the situation changes, as illustrated in Figure~\ref{fig-pinch}.
In the left plot, we have a torus with spacetime volume $V_4$ and time length $T$. It is pinched such that the space volume is  $\epsilon$ 
 during a time length $L$. 
The field $\phi(t)$ changes uniformly from 0 to $\delta/2$ over a distance $L/2$ in the lower red part. In the blue part  $\phi=\delta/2$, and in the upper red region, the field changes uniformly from $\delta/2$ to $\delta$. 
The total matter action of this field configuration is then 
\begin{equation}\label{jan50}
\Big( \frac{\delta}{L}\Big)^2 L \, \epsilon = \delta^2 \frac{\epsilon}{L},
\end{equation}
Clearly, this value can be made arbitrarily small when $\epsilon \to 0$. On the right plot, we also have a torus with spacetime volume $V_4$ and time length $T$. For this geometry, the matter action is minimal for a field changing uniformly from 0 to $\delta$, when we move from bottom to top; i.e., we obtain a matter action
\begin{equation}\label{jan51}
\Big( \frac{\delta}{T}\Big)^2 V_4  = \delta^2 
\frac{V_4}{T^2},
\end{equation}
which is bounded from below when $V_4$ and $T$ are fixed, as they are in our computer simulations. The conclusion is the same for a more careful analysis of the full minisuperspace action \rf{jax2}. Thus, we have a picture where for small $\delta$, the effect of the scalar field is small, and we can say that the scalar field couples to and follows the geometry. However, when $\delta > \delta_c$, a transition occurs: the scalar field pinches the geometry to a spatial volume which is small or maybe even zero, and all changes of  $\phi$ take place in this region of very small volume. Thus, $\phi$ basically separates a spacetime that has a non-trivial winding number in the time direction in two parts: {one (of cutoff size) with a winding number and  one} without a winding number.

\begin{figure}[t]

\centerline{\includegraphics[width=0.49\textwidth]{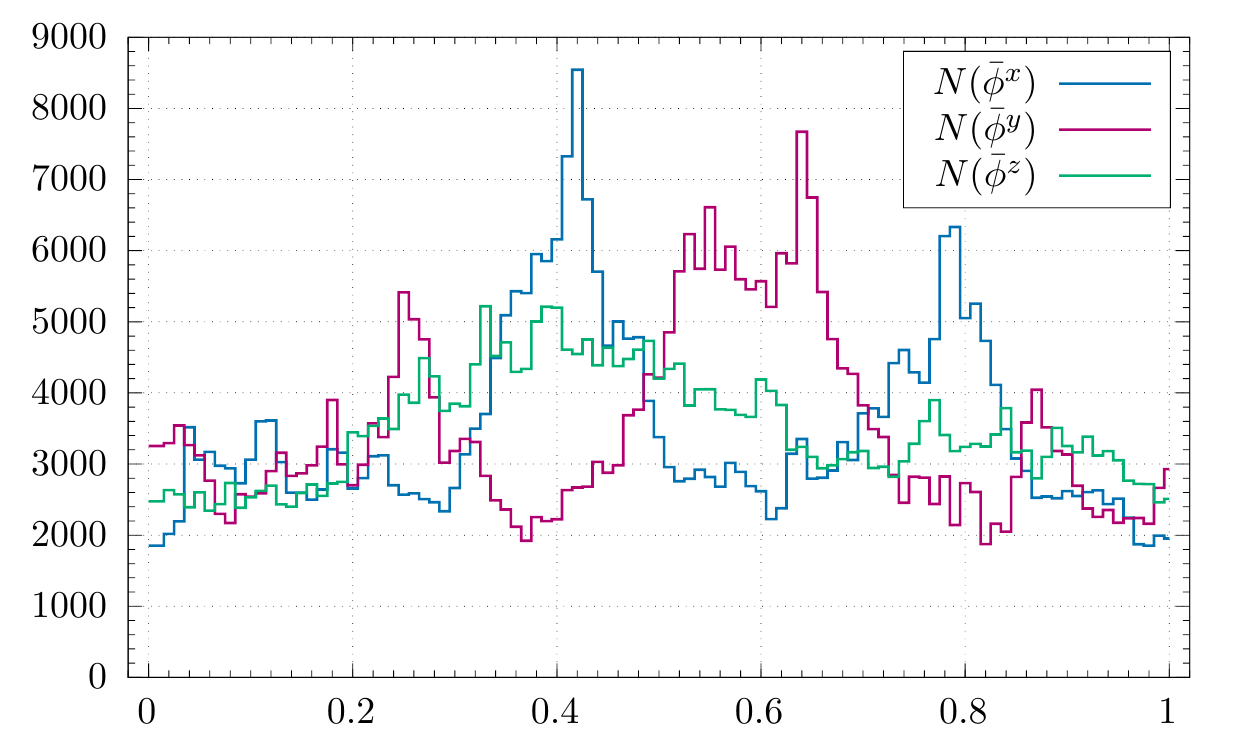}
\includegraphics[width=0.49\textwidth]{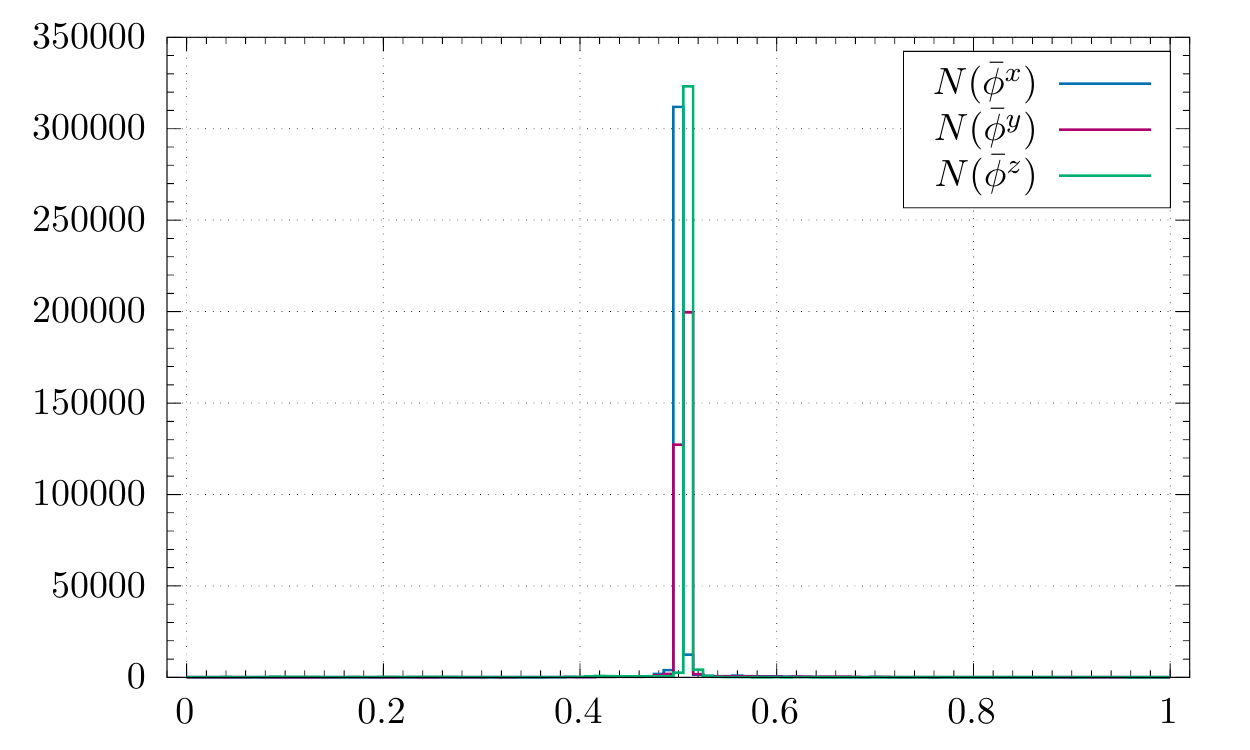}}
\caption{Left: the projection of four-volume, as defined by (\ref{hypers_vol_eq_psi})  in one spatial direction ($x$, $y$ or $z$) for a typical  CDT configuration  with  small jump magnitude $\delta=0.1$. Right: The projection of four-volume, as defined by (\ref{hypers_vol_eq_psi}),  in one spatial direction ($x$, $y$ or $z$) for a typical  CDT configuration  with  large jump magnitude $\delta=1.0$. The horizontal axis is $\bar{\phi}/\delta$.}
\label{1dsmall}
\end{figure}

This analysis is of course based on the very simple minisuperspace action, which might be a good description in the time direction, but not necessarily in the spatial directions. We have thus studied the situation when we have dynamical scalar fields with winding numbers in the spatial directions, using MC simulations. Thus, there is no minisuperspace approximation,  and it is the full quantum theory we consider. In Figure~\ref{1dsmall}, we show the volume profiles for a typical configuration for a small $\delta$ and a typical configuration for a large $\delta$. As coordinates in the spatial directions, we use the classical scalar fields $\psi$ with non-trivial winding numbers as described above. It is seen that the top and bottom part of Figure~  \ref{1dsmall} match very well with the right and left part of Figure~\ref{fig-pinch}. All the change of $\psi$ in the left part of Figure~  \ref{1dsmall} takes place in a a region where the spatial volume is very small, while $\psi$ is nearly constant where the volume is large.

To summarize: A scalar field with a winding number seems to have a dramatic effect on the geometry and can even induce a phase transition where the geometry is pinched or even becomes cut into pieces. {However, presently we do not have a simple
matter model, which in a natural way provides us with a such a scalar field.  }

\section{Discussion}\label{sec7}

In this review article we have summarized how the study of quantum universes with  toroidal topology can help us understand better the quantum nature of geometry at and close to the Planck scale. 

Firstly, the use of toroidal spatial topology allowed us to obtain a more complete CDT phase diagram. As discussed, the fact that topology
at all plays a role for the transitions is most likely a finite volume
effect and related to rearrangements of the geometric structures. 
For the size of volumes used in the computer simulations, some of these rearrangements may be easier implemented when the spatial topology is toroidal. Furthermore, the ease with which such changes take place probably depends on the Monte Carlo algorithms used, but presently we only have available the ones based on the so-called Pachner moves. 

Secondly, the picture of an emergent background geometry around which there are relatively small quantum fluctuations is corroborated by the results
obtained using toroidal spatial topology. This picture was first 
obtained when the spatial topology was that of $S^3$, but the emergent
background geometries are very different in the two cases.
It should be remembered that no background geometry 
is imposed in the CDT path integral, so it is non-trivial that two four-dimensional background geometries that are so different can emerge when  spherical and toroidal spatial topology are imposed. Likewise, it is 
non-trivial that  minisuperspace descriptions respecting 
the spatial topologies work so well in both cases. 
In the toroidal case, one is able to measure genuine quantum correction terms in the minisuperspace 
effective action. These terms can in principle be compared to correction
terms of minisuperspace models derived from a continuum theory of quantum gravity. 

Thirdly, the use of toroidal spatial geometry allows us to dissect in some detail global aspects of the geometries of typical path integral configurations by using the existence of non-contractible loops. The study of the lengths of the shortest non-contractible loops passing through  vertices on the dual lattice provides us with a picture of  
a typical spatial toroidal path integral geometry 
as a semiclassical toroidal ``core''
with large, quite fractal  outgrowths. By using a representation of the torus $T^3$ as a periodic structure in $\mathbb{R}^3$, this becomes particularly transparent. 

Finally, we can use the toroidal spatial 
topology to introduce spatial coordinates. Of course one nice aspect of 
the CDT formalism is that it is independent of spatial coordinates.
However, as discussed, coordinates {\it can} be useful and help us 
understand aspects of  the geometry, if chosen appropriately. We described two
ways to introduce coordinates, both taking advantage of the toroidal 
topology. The starting point for both coordinate systems was 
the definition of an elementary cell which defined the three-torus 
as a periodic structure in $\mathbb{R}^3$. With the first set of coordinates, which we denoted pseudo-Cartesian coordinates, we tried, loosely speaking,
to use the geodesic distance from the boundaries as coordinates. The
study of these coordinates revealed a lot about the fractal structure 
of the geometry involved, but the same fractal structure made 
the coordinate system less useful for other purposes since the hypersurfaces of a constant coordinate consisted of many disconnected parts. To deal with this problem, we introduced instead classical 
scalar fields with non-trivial boundary conditions at the 
boundaries of the elementary cell. Since these scalar fields are 
solutions to the Laplace equation for the given geometry in the interior of the elementary cell, it is ensured that they will not have local 
maxima or minima in the interior of the cell. In this way, they work well as coordinates for a given geometry, and we are presently trying to 
express a number of observables in terms of these coordinates. \mbox{Figure~\ref{phi-xy_plot}} shows the fractal structure of a geometry viewed with the help of these coordinates.

The scalar fields that we used to define the coordinates in these quantum universes were purely classical fields, not interacting with the quantum geometry. In the real world, we have matter fields interacting with geometry, and in particular, it has been speculated that scalar fields may play a very important role by creating the inflation that might have taken place in our Universe. From a technical point of view, it is easy to include scalar fields in the action such that they become dynamical fields interacting with the geometry. We have already done this in the case where no non-trivial topology was imposed on the scalar field. In that case, the effect of the scalar fields on the geometry was relatively minor and could be understood as equivalent to a change in the bare coupling constants of the model without  scalar fields. However, the situation seems different if the scalar field has a non-trivial winding number, and the results mentioned above indicate that there can be new phase transitions of the geometry associated with the presence of such scalar fields. Details of this are currently being explored, as is the search for matter models where such non-trivial winding numbers of the scalar fields can occur in a cosmological context.
    
\vspace{6pt}

\subsection*{Acknowledgments} {Z.D. acknowledges support from the National Science 
Centre, Poland, grant \\ 
no. 2019/32/T/ST2/00390. J.G.-S. acknowledges support of the grant no. \\
UMO-2016/23/ST2/00289 from the National Science Centre Poland. A.G. acknowledges support by the National Science Centre, Poland, under grant no. 2015/17/D/ST2/03479. J.J. acknowledges support from the National Science Centre, Poland, grant  
no. 2019/33/B/ST2/00589. D.N. acknowledges support from National Science Centre, Poland with grant no. 2019/32/T/ST2/00389.}


\end{document}